\documentclass[aps,twocolumn,preprintnumbers,prd,showpacs,showkeys,superscriptaddress,nobibnotes,floatfix,longbibliography,nofootinbib]{revtex4-1}
\pdfoutput=1
\usepackage[english]{babel}
\usepackage{amsmath,amssymb,amsbsy,amstext,amsthm,booktabs,exscale,relsize,slashed,graphicx,amsfonts,upgreek,xcolor,amsmath,setspace, ulem}
\usepackage{multirow,dcolumn,bm,enumerate,url,hyperref}
\usepackage[capitalise]{cleveref}
\usepackage{breakurl}
\usepackage{fontawesome}
\usepackage{xspace}
\definecolor{lightsabergreen}{rgb}{.14,.64,.14}
\definecolor{lightgreen}{rgb}{.14,.44,.14}
\usepackage{subfigure, tabularx}
\usepackage{comment}
\hypersetup{
  colorlinks   = true, 
  urlcolor     = lightsabergreen, 
  linkcolor    = lightsabergreen, 
  citecolor   = lightsabergreen 
}

\newcommand{\gev}{{\rm GeV}}
\newcommand{\mev}{{\rm MeV}}

\newcommand{\edits}[1]{\textcolor{black}{#1}}

\newcommand{\ra}[1]{\renewcommand{\arraystretch}{#1}}
\usepackage{printlen}
\begin{document}

\preprint{
\begin{minipage}{5cm}
\small
\flushright
SLAC-PUB-17743
\end{minipage}} 

\count\footins = 1000
\title{\bf Dark Matter-Induced Baryonic Feedback in Galaxies}

 \author{Javier F. Acevedo}
 \affiliation{\smaller Department of Physics, Engineering Physics, and Astronomy, Queen's University, Kingston, Ontario, K7L 2S8, Canada}
 \affiliation{\smaller Arthur B. McDonald Canadian Astroparticle Physics Research Institute, Kingston ON K7L 3N6, Canada}
 \affiliation{\smaller Particle Theory Group, SLAC National Accelerator Laboratory, Stanford, CA 94035, USA}
 
 \author{Haipeng An}
\affiliation{\smaller Department of Physics, Tsinghua University, Beijing 100084, China}
\affiliation{\smaller Center for High Energy Physics, Tsinghua University, Beijing 100084, China}
\affiliation{\smaller Center for High Energy Physics, Peking University, Beijing 100871, China}	
\affiliation{\smaller Frontier Science Center for Quantum Information, Beijing 100084, China}

  \author{Yilda Boukhtouchen}
 \affiliation{\smaller Department of Physics, Engineering Physics, and Astronomy, Queen's University, Kingston, Ontario, K7L 2S8, Canada}
 \affiliation{\smaller Arthur B. McDonald Canadian Astroparticle Physics Research Institute, Kingston ON K7L 3N6, Canada}

 \author{Joseph Bramante}
 \affiliation{\smaller Department of Physics, Engineering Physics, and Astronomy, Queen's University, Kingston, Ontario, K7L 2S8, Canada}
 \affiliation{\smaller Arthur B. McDonald Canadian Astroparticle Physics Research Institute, Kingston ON K7L 3N6, Canada}
 \affiliation{\smaller Perimeter Institute for Theoretical Physics, Waterloo, ON N2J 2W9, Canada}

\author{Mark L. A. Richardson}
 \affiliation{\smaller Department of Physics, Engineering Physics, and Astronomy, Queen's University, Kingston, Ontario, K7L 2S8, Canada}
 \affiliation{\smaller Arthur B. McDonald Canadian Astroparticle Physics Research Institute, Kingston ON K7L 3N6, Canada}

 \author{Lucy Sansom}
 \affiliation{\smaller Department of Physics, Engineering Physics, and Astronomy, Queen's University, Kingston, Ontario, K7L 2S8, Canada}
 \affiliation{\smaller Arthur B. McDonald Canadian Astroparticle Physics Research Institute, Kingston ON K7L 3N6, Canada}
 
\begin{abstract}

We demonstrate that non-gravitational interactions between dark matter and baryonic matter can affect structural properties of galaxies. Detailed galaxy simulations and analytic estimates demonstrate that dark matter which collects inside white dwarf stars and ignites Type Ia supernovae can substantially alter star formation, stellar feedback, and the halo density profile through a dark matter-induced baryonic feedback process, distinct from usual supernova feedback in galaxies.

\end{abstract}
\maketitle

\section{Introduction}

The nature of dark matter and of its non-gravitational interactions remains elusive, even as broad evidence for its existence accumulates through its gravitational dynamics in galaxies and its cosmological imprint on the early universe. In tandem with increasingly-sensitive direct detection experiments, dark matter's indirect effects on astrophysical environments can help reveal its properties. In this work we demonstrate that dark matter can alter the formation and morphology of galaxies through a Type Ia supernova-driven baryonic feedback process.

The cold, collisionless dark matter paradigm of $\Lambda$CDM cosmology, in which dark matter's influence on the Standard Model sector is gravitational, successfully models \edits{many} astrophysical and cosmological observations. \edits{These include cosmological fits to the primordial power spectrum \cite{Planck:2018vyg} and the dynamics of dark matter in galaxy clusters \cite{Vegetti:2023mgp, Tulin_2018}.} Cosmological simulations implementing the $\Lambda$CDM model, however, have raised questions about whether collisionless, non-interacting dark matter is consistent with observations of small-scale structure. \edits{$\Lambda$CDM N-body simulations predict steep, cuspy dark matter density distributions in dwarf galaxies, whereas observations show more diversity between cusped and cored inner density profiles in otherwise very morphologically similar galaxies. This problem, initially dubbed the ‘‘core-cusp" problem~\cite{Oh_2011}, is now considered to be a ‘‘diversity problem" \cite{Oman_2015}.} This implies that either the collisionless dark matter assumption ought to be revisited, or that structures at galactic scales are not solely determined by dark matter's gravitational effects, and baryonic effects are essential to a full picture. 

Increasingly-powerful computational \edits{codes} have enabled fluid dynamics at sub-galactic scales in cosmological simulations \edits{\cite{Springel_2005,Hopkins:2014qka, 2019ascl.soft09010S, Teyssier_2002, Schaller_2024, Chang_2017}. From these,}  the importance of modeling baryonic feedback effects has become evident. \edits{Sources of baryonic feedback include stellar feedback from supernovae, stellar winds and active galactic nuclei. For instance,} stellar feedback (or `baryonic' feedback) due to supernova explosions can heat gas locally and, in sufficient amounts, lead to gas outflow events that regulate star formation in smaller galaxies. Furthermore, these gas outflows, through the gravitational potential fluctuations they cause, can flatten the centre of dark matter density profiles. Indeed, past works have shown that explicit hydrodynamical modeling of gas cooling, star formation and supernova-induced gas outflows could \edits{produce cored density profiles in simulations} \cite{Governato_2012}. In the same vein, hydrodynamical simulations indicate baryonic feedback, by quenching star formation, could result in fainter and less observable satellite galaxies around large galaxies like the Milky Way \cite{Koposov_2009,Kim:2017iwr}. 

Though baryonic feedback effects are Standard Model in nature, as their name suggests, here we find that dark matter may play a role in the amount and strength of these effects. Past work has shown that \edits{asymmetric} dark matter may explain observed Type Ia supernovae, which have been traced to sub-Chandrasekhar white dwarfs, by igniting white dwarf cores \cite{Bramante:2015cua,Acevedo:2019gre,Janish:2019nkk,Bramante:2023djs}. \edits{Asymmetric models of dark matter are motivated by the fact that the dark matter and visible matter densities today are within an order of magnitude of each other ($\Omega_{DM} \approx 5 \Omega_{B}$). These models propose that this is due to both having a matter-antimatter asymmetry generated in the very early universe by some shared mechanism that has since frozen-out \cite{Petraki:2013wwa,Zurek:2013wia}. In the context of dark matter ignition of white dwarfs, the absence of dark matter self-annihilations in the white dwarf leads to a faster rate of dark matter accumulation in compact objects, which leads to more efficient white dwarf ignition.} Dark matter ignition of white dwarf explosions leads to an additional source of baryonic feedback, which, while sometimes energetically subdominant compared to standard baryonic feedback, has qualitatively different properties that make its effect on galaxy evolution interesting to study. Dark matter-induced baryonic feedback, or dark baryonic feedback for short, has an additional dependence on the dark matter density in the neighbourhood of white dwarfs, leading to explosions preferentially in the center of the galaxy, where the dark matter density is highest. Furthermore, dark baryonic feedback can be expected to shut itself off in certain cases -- both through suppression of star formation and thinning of the central dark matter density.

In this Article, we explore the effects of dark baryonic feedback on star formation and dark matter density in a small galaxy. In Section II, we give an overview of dark matter ignition of Type Ia supernovae. \edits{In Section III, we review the mechanics of baryonic feedback and the gravitational potential fluctuations it induces. In Section IV, we outline the numerical implementation of dark baryonic feedback, and discuss simulation results in Section V. In Section VI, we conclude}.

\section{Dark matter ignition of Type Ia supernovae}

Several works have demonstrated that dark matter could ignite Type-Ia supernovae, including those proceeding from sub-Chandrasekhar white dwarf progenitors. The Type-Ia progenitor problem is a longstanding area of astrophysical interest \cite{Maoz:2013hna}, heightened by the observation of sub-Chandrasekhar Type-Ias \cite{10.1093/mnras/stu350}. Proposed dark matter Type-Ia supernova ignition channels include formation and collapse of asymmetric dark matter cores \cite{Bramante:2015cua,Acevedo:2019gre,Janish:2019nkk}, evaporation to Hawking radiation of light black holes formed from the collapsed dark matter \cite{Acevedo:2019gre,Janish:2019nkk}, heavy dark matter annihilation or decay to Standard Model particles \cite{Graham:2018efk}, pycnonuclear reactions catalyzed by the capture of charged massive particles \cite{Fedderke:2019jur}, as well as the transit of primordial black holes \cite{Graham:2015apa,Montero-Camacho:2019jte,Steigerwald:2021vgi} or massive composite dark matter states \cite{Acevedo:2020avd,Acevedo:2021kly,Raj:2023azx}. Unlike other ignition mechanisms, such as accretion of stellar material from a neighboring companion \cite{1982ApJ...253..798N,Han:2003uj,Wang:2018pac}, binary mergers \cite{doi:10.1111/j.1365-2966.2011.19361.x,doi:10.1093/mnras/stw1575}, or helium-shell ignition \cite{2041-8205-770-1-L8}, these dark matter ignition channels do not require the presence of a binary companion. More importantly, massive white dwarfs need not reach the Chandrasekhar threshold to explode. Intriguingly, dark matter ignition of Type-Ia supernovae may explain recent findings indicating that a significant number of these events proceed from sub-Chandrasekhar mass white dwarfs \cite{Scalzo:2014wxa,NearbySupernovaFactory:2014mcg,Bramante:2015cua,Steigerwald:2021bro} in single stellar systems \cite{2015Natur.521..332O,Maoz:2013hna} and the observed spatial distribution of Ca-rich transients in dwarf galaxies \cite{Smirnov:2022zip}.

While all dark matter ignition mechanisms could produce dark baryonic feedback, to illustrate this new effect we focus on \edits{Type-Ia} ignition through the gravitational collapse of asymmetric dark matter \cite{Bramante:2015cua,Acevedo:2019gre,Janish:2019nkk,Steigerwald:2022pjo}. In this ignition channel, asymmetric dark matter particles are captured and thermalized in a small region of the white dwarf core. \edits{The dark matter's asymmetry allows sufficient dark matter particles to be captured and thermalized within a white dwarf lifetime.} The dark matter accumulated at the core grows in mass over time, eventually becoming self-gravitating and collapsing under its own weight. As collapse proceeds, the virialized dark matter particles shed their potential energy by repeatedly scattering against white dwarf nuclei, heating a small central region to the critical temperature needed for thermonuclear runaway. This ignition channel produces sharp predictions in our galaxy simulations, because it makes a sharp prediction for ignition time, as compared with $e.g.$ stochastic white dwarf ignition through encounters with primordial black holes and other dark compact objects \cite{Montero-Camacho:2019jte,Acevedo:2020avd,Acevedo:2021kly,Steigerwald:2021vgi,Raj:2023azx}. 

The strength of the dark baryonic feedback is regulated by a newly-formed white dwarf’s explosion time. This is determined by the rate at which dark matter accumulates, thermalizes and eventually collapses inside the white dwarf. Each of these timescales along with the white dwarf ignition process are detailed in Appendix \ref{app:ignition}. In practice, for specific dark matter parameters, only one of the above timescales is the bottleneck to the ignition process. In the case of parameters used in our simulations, the accumulation rate is the relevant bottleneck, and scales linearly with dark matter density $\rho_X$, and inverse-linearly with the dark matter velocity $v^{-1}_{X}$. For this reason, white dwarfs explode more from dark matter ignition in the center of galaxies.

\section{Mechanism for dark baryonic feedback}

We now outline the dynamics of baryonic feedback and discuss how gravitational potential fluctuations due to dark matter-induced baryonic feedback will affect galactic halos. We use a similar approach as \cite{Pontzen:2011ty}, and consider distinct features of dark matter-induced feedback. For the moment we assume a spherically symmetric, power-law gravitational potential of the form $V(r, t) = V_0(t) r^n$. The virial theorem relates the average potential energy of a particle $V$ and its total energy $E_0$, $\langle V \rangle = \frac{2 E_0}{2 + n}$.  If the gravitational potential undergoes a sudden (instantaneous) change by $\Delta V_0$, such as that induced by a supernova explosion, the particle's total energy changes by $\Delta E_1$, where
$
    \Delta E_1 = \frac{2 E_0}{2 + n} \frac{\Delta V}{V_0}.
$

When modeling orbits of individual particles, we can assume a central potential sourced by a nearly constant density, i.e. $n = 2$, corresponding to a harmonic oscillator potential. The orbit of the dark matter particle is given by $r(t) = A \cos (\omega t + \psi)$ and its evolution after sudden shifts in $V$ is modelled as in \cite{Pontzen:2011ty}. Some time after energy injection, it is assumed that the blown out gas cools and recollapses so that the gravitational potential is nearly restored to its original form. As we will see in simulations, there are qualitative differences between standard vs.~dark matter-induced baryonic feedback, evident in Figures~\ref{fig:analyticandsim} and \ref{fig:galdistros}. 
\begin{figure*}[t!]
    \centering
    \includegraphics[width=\linewidth]{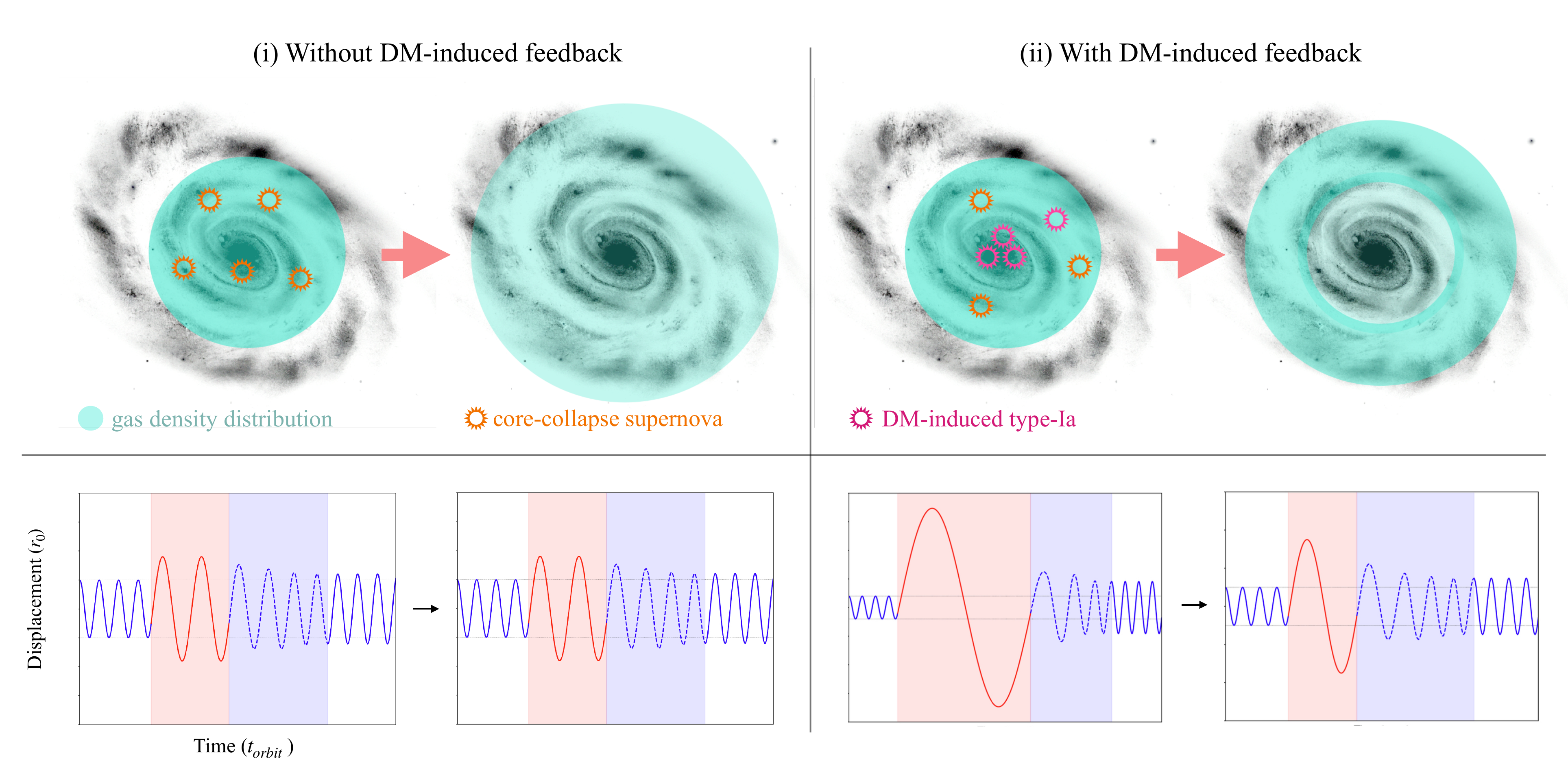}
    \caption{Schematic of baryonic feedback in a galaxy, with and without DM-induced Type Ia feedback. {\em Top Left:} Baryonic feedback from core-collapse (‘‘CC") supernovae and star formation blows out gas uniformly in regions of star formation. {\em Top Right:} DM-induced Type Ia feedback blows out preferentially in the galaxy's center, where DM is denser and rapidly ignites Type Ia explosions. {\em Bottom Left:} Subsequent baryonic feedback epochs cause similar shifts in test particle orbits. {\em Bottom Right:} DM-induced baryonic feedback can shut itself off over time, since prior feedback reduces the central DM density and star formation.}
    \label{fig:AnalyticOrbit}
\end{figure*}
In particular, because dark baryonic feedback decreases the galactic center dark matter density and star formation rate, it decreases future supernova rates and also dark baryonic feedback itself. 

Figure \ref{fig:AnalyticOrbit} shows a qualitative comparison of standard and dark baryonic feedback in a galaxy. The bottom panels compare the evolution of a test particle orbit amplitude $A$ and phase $\psi$, after in $V \rightarrow V + \Delta V$, in semi-analytic modelling representative of the two feedback scenarios. The main differences between standard baryonic feedback and dark matter-induced baryonic feedback are as follows: (1) In standard baryonic feedback core collapse supernovae will be distributed throughout the galaxy wherever there are regions of star formation. In dark matter-induced baryonic feedback, dark matter-induced Type Ia explosions will occur most (and most rapidly) at the center of the galaxy, where the dark matter density is highest, and so dark matter accumulates in white dwarfs most rapidly in this region. (2) Baryonic feedback (of any origin) tends to decrease the density of dark matter at the center of galaxies. In the case of dark matter-induced feedback, this will decrease future dark matter-induced feedback, since white dwarfs will explode less and over longer timescales from dark matter accumulation. In the case of standard baryonic feedback, central dark matter density should not change the expected amount of feedback from star formation processes.

\begin{figure}[h!]
    \centering
    \includegraphics[width=.99\linewidth]{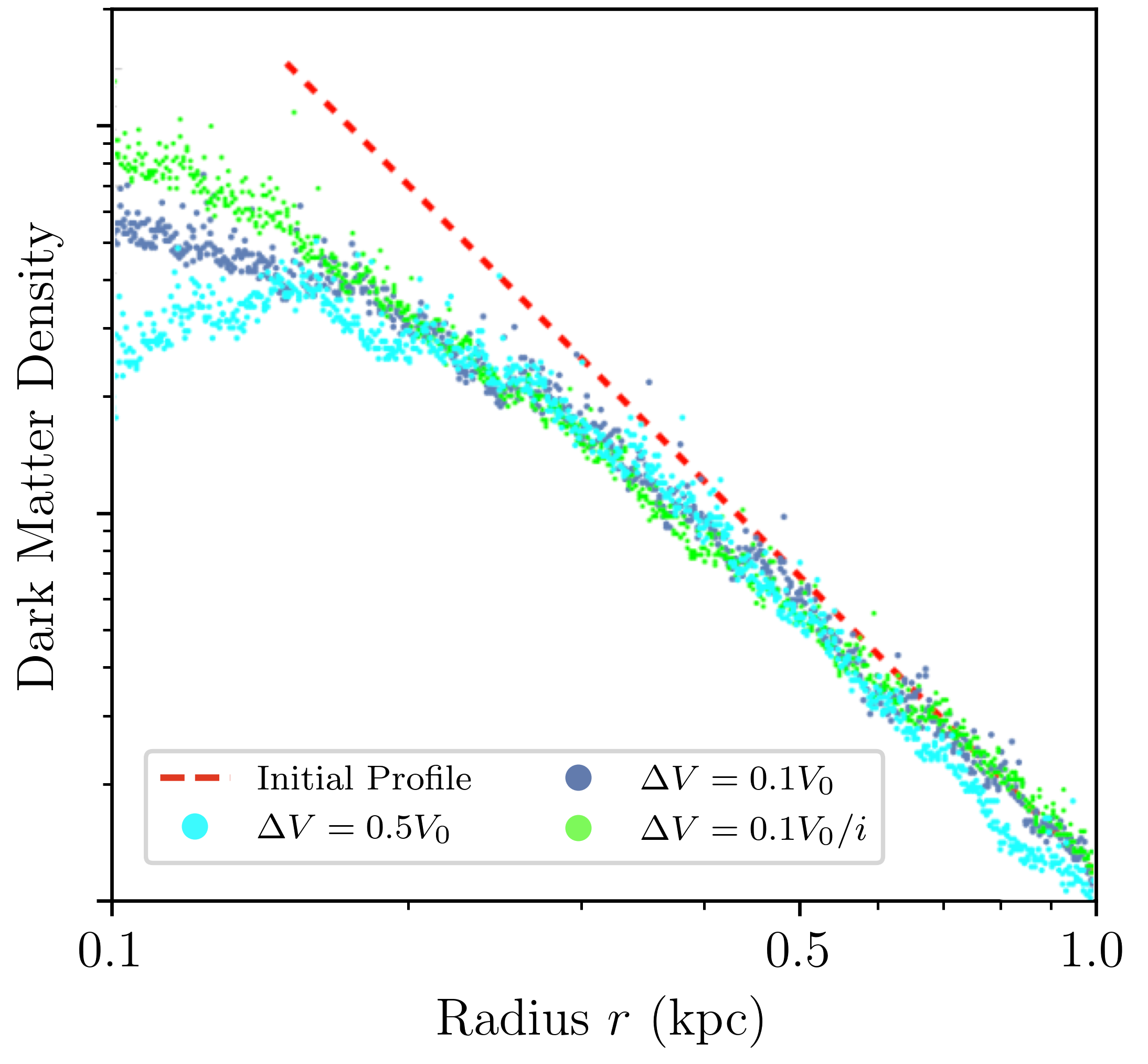}
    \caption{Semi-analytic results illustrating one aspect of dark baryonic feedback, comparing how test particles widen their orbits after eight shifts in central gravitational potential $\Delta V$, followed by a restoration of the original central potential. The case where shifts decrease over time ($\Delta V = 0.1 V_0/i$ ) illustrates a novel feature of dark baryonic feedback, that a flatter central potential decreases dark baryonic feedback over time.}
    \label{fig:semianalytic}
\end{figure}

In Figure \ref{fig:semianalytic} we show some simplified exploration of baryonic feedback in a semi-analytic model galaxy, subject to certain limits. We caution the reader that this semi-analytic treatment does not model the dynamical interplay of baryonic feedback, star formation, and dark matter-indicued feedback from white dwarfs exploding; for this we require a full simulation undertaken in Section \ref{sec:simulations}. Here we simply look at how different shifts to the central gravitational potential of the galaxy changes the distribution of test particles.
After an epoch of baryonic feedback decreases the central gravitational potential of a galaxy, whether that central potential of the galaxy increases to its original value through gas re-cooling will depend on the intensity of the feedback. In this work, we will be interested in a gradual gas recollapse process in which the central gravitational potential is changed over a timescale of $\sim 50$ Myr. We note that one of the variables governing the explosion time of a white dwarf due to dark matter capture is the white dwarf mass, with higher-mass white dwarfs exploding in a shorter time \cite{Bramante:2015cua,Acevedo:2019gre}. A single epoch of star formation therefore leads to multiple successive supernova events at different timescales. 

To provide some approximate modeling of late-stage dark baryonic feedback processes in a galaxy, we consider a population of dark matter particles, distributed according to a simplified halo profile near a galactic centre. For simplicity we use a power-law gravitational potential in the interior region given by $\rho(r) = \rho_0 r^{-2}$, to illustrate baryonic feedback effects in a regime with a cuspy internal profile. We use a code that explicitly calculates the position and acceleration of the dark matter particles to determine their orbits following changes in the gravitational potential, which we model as
\begin{equation}
    \Delta V(t) = - \kappa_i V_0  \left( 1- \left( 1 - \exp\left[ -\frac{t}{t_{\rm recollapse}} \right] \right) \right)
\end{equation}
where $\kappa_i$ is the factor by which the potential is decreased after a supernova explosion. The exponential term implements the gradual recollapse of the potential after the explosion time $t=0$, for a recollapse time $t_{\rm recollapse} \approx 50$ Myr, which we assume is much shorter than the time between explosions.

To illustrate a qualitative difference between standard vs.~dark matter-induced baryonic feedback, in Figure~\ref{fig:semianalytic} we show how assuming a fixed values for $\kappa_i$($=0.1,0.5$) differs from assuming that $\kappa_i$ decreases with each explosive epoch as $\kappa_i=\kappa_1/i$, for epochs $i=1,2,...8$. This mimics an effect of dark baryonic feedback we will find in Section \ref{sec:simulations}, which will tend to decrease the galactic center dark matter density and star formation rate, thereby decreasing future supernova rates and also dark baryonic feedback itself.

\section{Simulation of Dark Baryonic Feedback}
\label{sec:simulations}

In this section, we describe the numerical implementation of dark matter-induced white dwarf explosions in an isolated dwarf galaxy simulation.

\subsection{Galaxy initial conditions}

The initial conditions for the isolated galaxy were constructed using \textsc{MakeNewDisk}, a code which sets up galactic components in quasi-equilibrium based on the GADGET framework \cite{Springel05, Springel_2021}. The galaxy properties are summarized in Table \ref{tab:Init}. 

\begin{table*}[ht!]
    \centering
    \ra{1.3}
    \begin{tabular}{@{}lllllll@{}}
    \toprule[0.08em]
    \multicolumn{2}{l}{\textbf{Galaxy Properties}} & & & \multicolumn{2}{l}{\textbf{Component mass fraction}} & \\
    \midrule[0.05em]
     & Total mass  & $1.9 \times 10^9 \ M_{\odot}$ & & & Dark matter & 0.9615 \\
     & Halo concentration     &  10 & & & Gas disk & 0.0175 \\
     & Spin parameter & 0.04 & & & Stellar disk & 0.0175\\
     & Particle Mass & $3 \times 10^3 \ M_\odot$ & & & Stellar bulge & 0.0175\\
     \bottomrule[0.08em]
    \end{tabular}
    \caption{Isolated galaxy parameters used for creating initial conditions in \textsc{MakeNewDisk}. \edits{We note that the particle mass given here is referring to both dark matter and baryonic particles in the simulation.}}
    \label{tab:Init}
\end{table*}

\subsection{Isolated galaxy evolution in GIZMO}

\begin{table*}[t!]
    \centering
    \ra{1.3}
    \begin{tabularx}{0.95\textwidth}{@{}llX@{}}
    \toprule[0.08em]
    Gravity & \texttt{TreePM} & Tree method for short-range calculations;  particle-mesh grid solver at long ranges. \cite{Bagla_2002} \\
    \midrule[0.05em]
    Hydrodynamics & \texttt{HYDRO\_PRESSURE\_SPH} & “Pressure" or “density-independent" smoothed-particle hydrodynamics solver. \cite{Hopkins_2012} \\
    & \texttt{COOLING} & Radiative cooling and heating. \cite{Hopkins_2018_cooling, cooling_uv}\\
    \midrule[0.05em]
    Star formation & \texttt{GALSF} & Star formation: modeling the conversion of gas particles into star particles. \cite{Springel_2003}\\
     & \texttt{GALSF\_SFR\_CRITERION = (0+1)} & Sets a density threshold, as well as a local self-gravitation requirement, for star formation. \cite{Hopkins_2013, Hopkins_2018}\\
     \midrule[0.05em]
    Stellar feedback & \texttt{GALSF\_FB\_MECHANICAL } & Explicit mechanical feedback coupling algorithm, which is shown to accurately resolve energy and momentum conservation during energy injection from supernova events. \cite{Hopkins_2014, Hopkins_2018, Kimm_2014, 10.1093/mnras/stv562, 10.1093/mnras/stw3034}\\
     & AGORA project core-collapse model & Determines the number of core-collapse supernovae occurring in a star particle, using IMF-averaged values. \\
     & & Supernovae all occur when the particle reaches 5 Myr of age, and each inject $10^{51}$ erg of energy into their surroundings. \cite{2016ApJ...833..202K} \\
    \bottomrule[0.08em]
    \end{tabularx}
    \caption{List of modules and settings used in GIZMO simulation of isolated galaxy.}
    \label{tab:GIZMO}
\end{table*}

We evolve this galaxy for $2.5$ Gyr using the GIZMO codebase, for different baryonic feedback mechanisms. GIZMO is a multi-purpose fluid dynamics and gravity code \cite{Hopkins_2015} descended from the cosmological simulation code GADGET-2 \cite{Springel_2005}. It supports a variety of physical processes -- notably several hydrodynamics codes, star formation and stellar feedback. We \edits{determine the supernova rate and energy injection} for the standard, core-collapse supernova feedback case, using the AGORA model, which assumes an averaged number of events set by an initial mass function (IMF) describing the mass distribution of main sequence stars within each star particle, and occurring at $5$ Myr. \edits{We then use the mechanical feedback algorithm in GIZMO to inject the appropriate energy to neighbouring particles.} The modules and settings used in this paper are further detailed in Table \ref{tab:GIZMO}.

\subsection{Implementation of white dwarf formation}

Relative to prior simulations, we need to explicitly track the formation and explosion of white dwarf stars. We describe our methodology below. 

\begin{figure}[h!]
    \centering\includegraphics[width=0.5\textwidth]{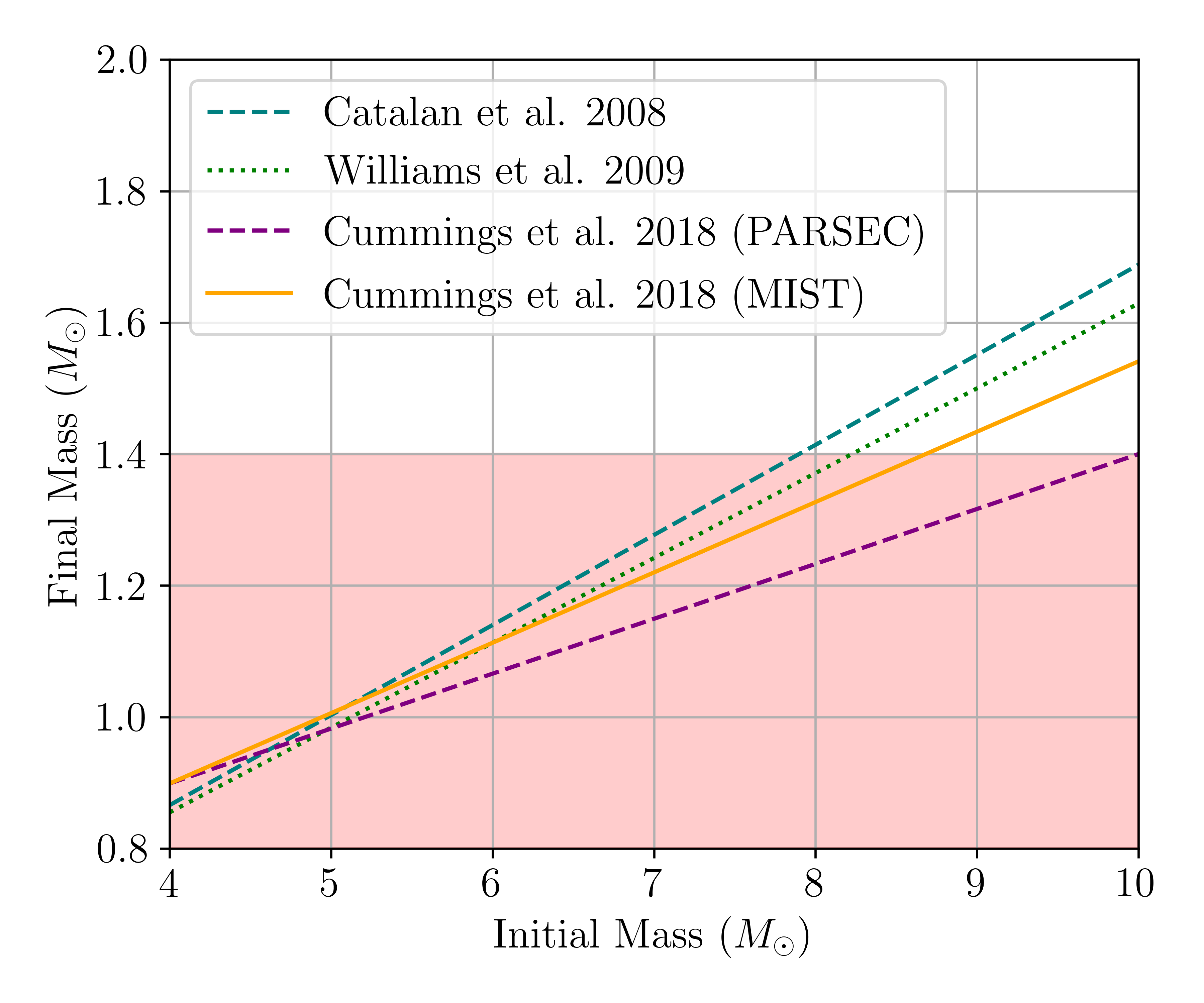}
    \caption{Comparison of the initial-final mass relation (IFMR), relating the mass of a white dwarf and the mass of its main sequence progenitor, given by four semi-empirical models. The PARSEC model has been extrapolated for initial masses beyond $8.2 \ M_\odot$, and the MIST model was extrapolated for masses beyond $7.2 \ M_\odot$. The shaded red region indicates white dwarfs which are candidates for dark baryonic feedback, and the white dwarfs of interest in this work.}
    \label{fig:IFMR}
\end{figure}

Low-mass main sequence stars, upon exhausting their helium reserves, blow out their outer layers and leave behind their stellar core, which lives on as a white dwarf. The mass of this white dwarf is related to the mass of its main sequence progenitor by an initial-final mass relation (IFMR), which can be determined from theoretical modeling \cite{Choi_2016}, or semi-empirically from analysis of white dwarf populations \cite{Catalan2008, Williams_2009, Cummings2018}. Semi-empirical analysis requires the choice of an isochrone model, such as the two implemented in the stellar evolutions PARSEC and MIST. Figure \ref{fig:IFMR} compares four IFMRs derived across three semi-empirical analyses. 
In this work, we chose to use the IFMR from the Cummings et al. PARSEC  model \cite{Cummings2018}: 
\begin{equation}
     M_{f} = (0.0835 \pm 0.0144) M_i + (0.565 \pm 0.073) M_{\odot}
     \label{eq:parsec}
\end{equation}
This relation was chosen because the observational data used in the Cummings et al. analysis had a high signal-to-noise ratio and low scatter compared to the other analyses.

We model main sequence stellar populations in each star particle using the initial mass function (IMF), a power law which describes the distribution of the masses of main sequence stars formed in a star-forming region. The probability of forming main sequence stars in each mass range was determined by integrating the IMF. Though various forms of the IMF have been proposed since its initial description in 1955, we choose to use the Chabrier IMF, 
\begin{equation}
\xi (m) = \left( \frac{3.0}{\mathrm{pc}^{3} M_\odot}\right)  m ^{-3.3} \mathrm{exp}\left[- \left(\frac{716.4 M_\odot}{m}\right)^{0.25} \right]
\end{equation}

\begin{figure}[h!]
    \centering
    \includegraphics[width=0.5\textwidth]{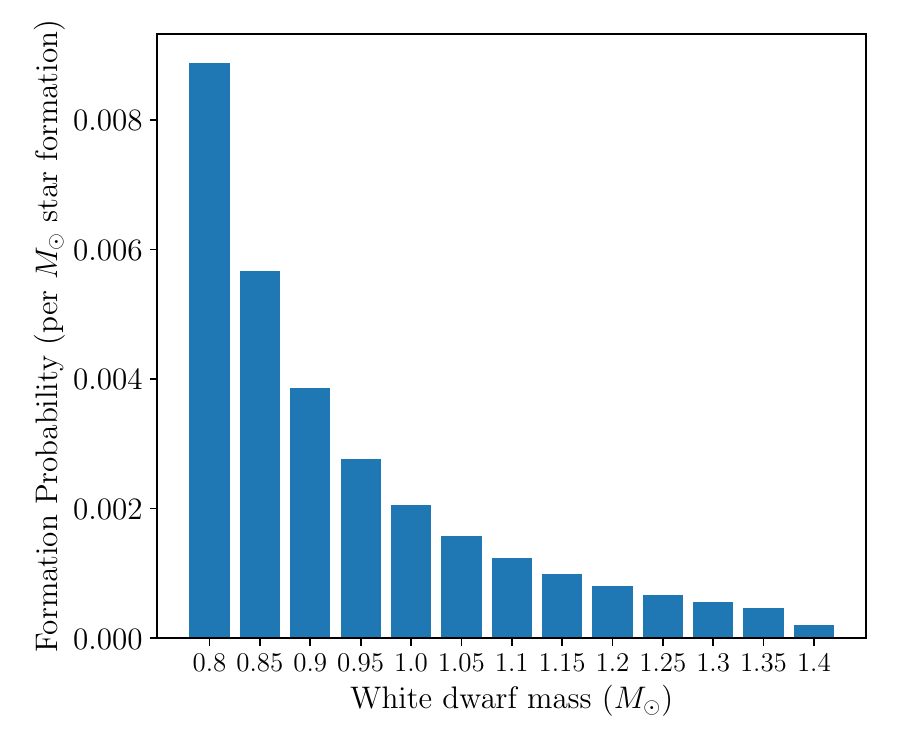}
    \caption{Probability of forming white dwarfs with in  between $0.8 M_\odot$ and $1.4 M_\odot$, per unit of star formation (in $M_\odot$). White dwarf masses are binned in mass ranges of $0.5 M_\odot$ in width.}
    \label{fig:WDProbabilities}
\end{figure}

In a star-forming region, the distribution of stars formed within the range [$M_i, M_f$] is then :
\begin{equation}
    N = \int_{M_i}^{M_f} \zeta(m) dm
    \label{eq:prob}
\end{equation}
where $\zeta(m) \propto \xi(m)$ is the probability distribution function obtained by normalizing the IMF over the mass range of main sequence stars ($\sim 0.08-100 M_\odot$). We are interested in determining the probability of forming white dwarf progenitor stars which will yield white dwarfs of masses $0.8 - 1.4M_\odot$, and do so using the chosen IMF and IFMR. The probability of forming a white dwarf in mass bins of width $0.5 M_\odot$ in this mass range of interest is plotted in Figure \ref{fig:WDProbabilities}.

We then determine the time taken for these white dwarf populations to form. The time after which a low-mass star leaves the main sequence and becomes a white dwarf depends on factors such as mass, metallicity, and rotation. To keep the modeling simple, we have used stars of solar metallicities and no rotation, as these factors have little effect on dark matter capture. With this assumption, the white dwarf formation time is determined from the main sequence progenitor lifetime, which is estimated using the Sun's evolutionary model as a benchmark, and follows the relation \cite{Hansen1999-za}
\begin{equation}
    t_{\rm MS} \simeq 10^{10} \  {\rm {yr}} \left(\frac{M_{\rm star}}{M_{\odot}}\right)^{-2.5}~.
\end{equation}
When a star particle is formed within the simulation, we use the formation probability described above to determine how many white dwarfs of each mass it will contain and save this in the particle data, along with the main sequence lifetime of their respective progenitor stars, $t_{MS}$.

\subsection{Implementation of dark baryonic feedback}

The explosion time for sub-Chandrasekhar white dwarfs in a dark-driven ignition scenario depends on:

\begin{itemize}
    \item \textit{The dark matter mass and dark matter-nucleon cross-section}, $m_X$ and $\sigma_{nX}$. These dark matter model parameters are set as global constants in the simulation.
    \item \textit{The white dwarf central density}, $\rho_{WD}$. We use the relation between white dwarf mass and central density obtained in Ref. \cite{Fedderke_2020} by numerically solving the Tolman-Oppenheimer-Volkov equation,

    \begin{equation}
        \rho_C(M_{WD}) \sim 1.95 \times 10^6 \mathrm{g /cm}^3 \left \lbrack \alpha(M_{WD})^{-2} - 1\right\rbrack^{3/2} \\
    \end{equation}

    where $\alpha(x) \simeq 1.0033 - 0.3087 x - 1.1652 x^2 + 2.0211 x^3 - 2.0604 x^4 + 
 1.1687 x^5 - 0.2810 x^6$.
    
    \item \textit{The local dark matter halo density and speed}, $\rho_X$ and $v_X$, in the neighbourhood of the white dwarf. As already mentioned, galaxies are initialized using \textsc{MakeNewDisk} standard initial conditions given in Table \ref{tab:Init}, which sets the initial distribution for dark matter shown in Figure \ref{fig:analyticandsim}. However, since baryonic feedback will shift this density profile during the simulation, it is necessary to update the dark matter density and velocity, since these quantities determine the DM-induced Type Ia rate for each star particle at each timestep. Hence, our code calculates these quantities 
    at each timestep by summing over all dark matter particles in radial bins, to obtain the number density of dark matter particles as a function of galactocentric radius $r$. The dark matter density and velocity in the neighbourhood of a each star particle at each timestep is then computed using this distribution. 
        
\end{itemize}

The white dwarf explosion time for each white dwarf mass is calculated once the star particle's age exceeds $t_{MS}$, using the formula for the ignition time, $t_{\rm crit} = M_{\rm crit}/C_X$ \edits{, where $M_{crit}$ is the critical dark matter mass that must be captured before the dark matter core collapses, and $C_X$ is the dark matter capture rate. See Appendix \ref{app:ignition} for a detailed account of these quantities.} These values are also stored in the particle data. At each simulation timestep, the age of the particle is checked against them to determine how many white dwarf supernovae are set off (if any) within the star particle. We assume that each Type Ia explosion injects $10^{51}$ erg of energy into its surroundings, equal to the energy injection from a core-collapse event \cite{2016ApJ...833..202K}. A standard Type Ia supernova from a Chandrasekhar-mass progenitor releases about 1.3 $\times 10^{51}$ ergs of energy, a quantity determined by the nuclear binding energy of the star, and thus its mass and carbon and oxygen content \cite{HOFLICH2006579}. We assume that, as these white dwarfs are in a similar mass range, the energy injected is also approximately equal. We use GIZMO's mechanical feedback formulation 
to inject the expected number of Type Ia supernovae once the explosion time had elapsed. A single star formation event can result in a string of Type Ia supernova events, as more massive white dwarfs explode sooner than their less-massive counterparts.

\section{Discussion}

\begin{figure*}[th!]
    \centering
    \includegraphics[width=\linewidth]{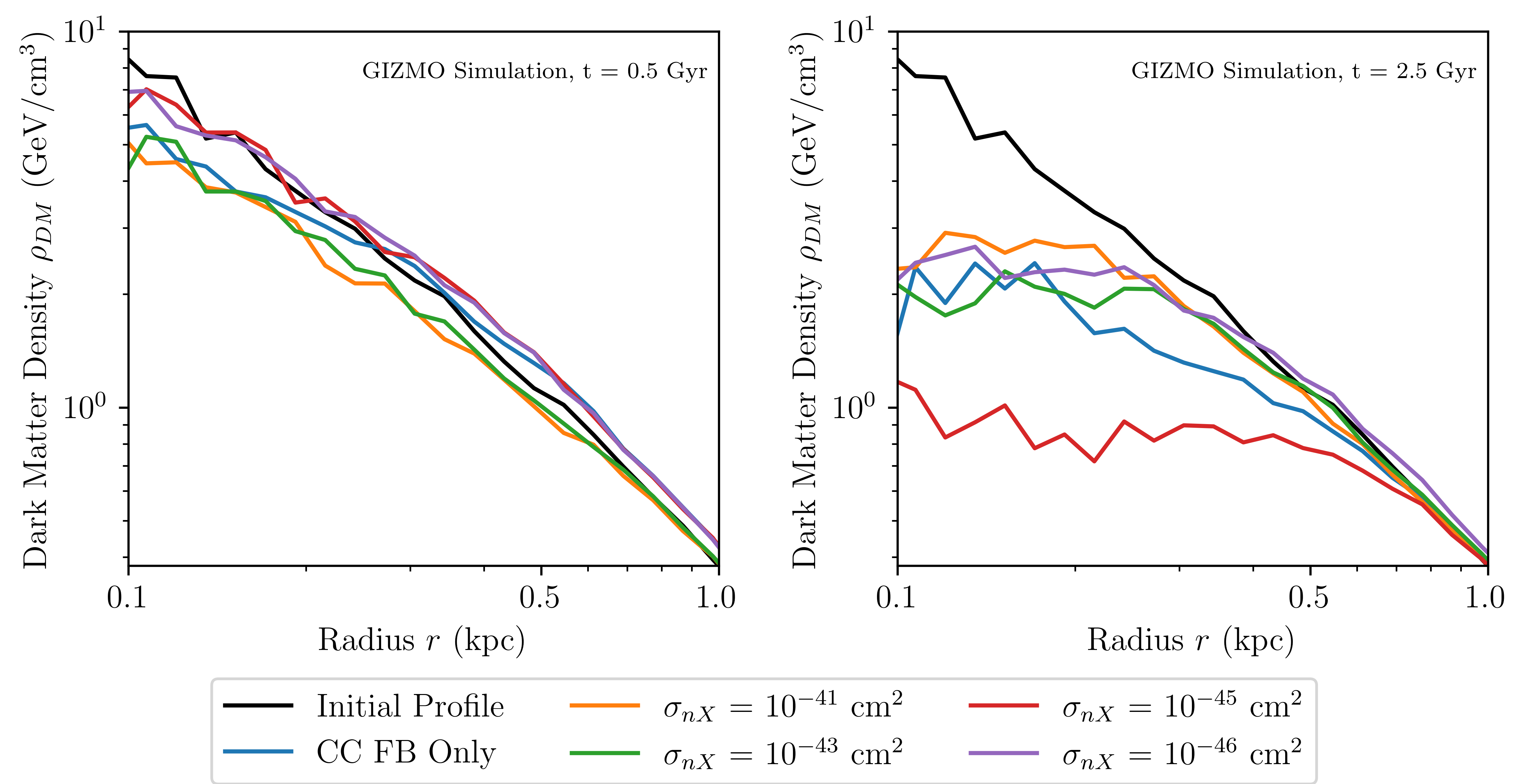}
    \caption{Resulting dark matter density profiles for different amounts of dark matter-induced baryonic feedback, at early ($t=0.5$ Gyr) and late ($t=2.5$ Gyr) simulation times, and for dark matter mass $m_X =10^7$ GeV and nucleon cross-section $\sigma_{nX}$ as indicated, where larger $\sigma_{nX}$ will cause more and prompter dark matter-induced white dwarf explosions.}
    \label{fig:analyticandsim}
\end{figure*}

In this section, we show our results from simulating dark baryonic feedback in our isolated dwarf galaxy for 2.5 Gyr, in addition to the usual core-collapse supernova feedback. 

The dark baryonic feedback simulations use a dark matter mass $10^7$ GeV and dark matter-nucleon cross-sections ranging from $10^{-41}$ to $10^{-46}$ cm$^2$. \edits{These values were selected because }this region of parameter space results in expected white dwarf explosion times of $\sim$ 10 Myr ($1$ Gyr) near the center of the simulated galaxy, for high (low) cross-sections\edits{, thus producing effects that can be explored within a $2.5$ Gyr simulation time. Much of this region of parameter space is also currently unconstrained by direct detection, $e.g.$ by experiments like Xenon-1T\cite{Aprile_2017}, which sets a bound of order $\sigma_{nX} \gtrsim 10^{-42}~{\rm cm^2} (m_X/10^{7}~{\rm GeV})$.} Results from these simulations can be seen in Figure \ref{fig:analyticandsim}.

\begin{figure*}[t!]
    \centering
    \includegraphics[width=\linewidth]{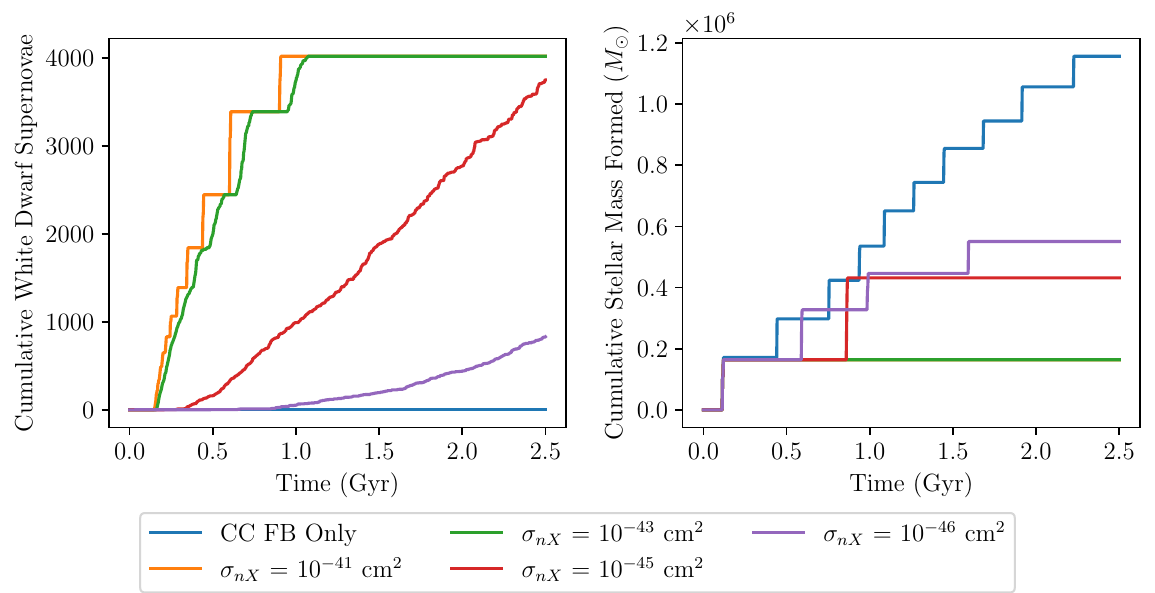}
    \caption{Evolution of Type Ia supernovae and stellar mass formed in $\sim 10^9~M_\odot$ galaxy simulations. {\em Left panel:} Number of dark matter-induced white dwarf supernovae over time, as a function of dark matter-nucleon scattering cross-section $\sigma_{nX}$ (for $m_X = 10^7$ GeV). {\em Right panel:} Same as left panel, but for stellar mass formed. We note that the orange and green curves are superimposed.}
    \label{fig:galdistros}
\end{figure*}

Figure \ref{fig:galdistros} shows the cumulative dark matter-induced white dwarf supernova rate across these simulations. At the highest dark matter-nucleus cross-sections, capture occurs very quickly, and so the white dwarf explosion time is \edits{dominated by its progenitor's} main sequence lifetime. We therefore see a very step-like behaviour in the Type Ia supernova count \edits{due to the mass binning we used in sampling the main sequence progenitor population, as we calculate main sequence lifetimes directly from stellar mass.} As we decrease the cross-section, the capture time increases, and white dwarfs explode at regular intervals (Figure \ref{fig:galdistros} $\sigma_{nX} = 10^{-45}, 10^{-46}$~cm$^2$), instead of exploding all together in bursts ($\sigma_{nX} = 10^{-41}, 10^{-43}$~cm$^2$). At sufficiently low cross-sections, the white dwarf explosion timescale exceeds the simulation time and we recover a galaxy with mostly core-collapse-driven feedback.  The abrupt cutoffs in dark matter-induced supernovae at $\sim 1$ Gyr in the highest $\sigma_{nX}$ simulations are due to an intense early period of baryonic feedback inhibiting galactic star formation completely. 

Figure \ref{fig:galdistros} also shows the cumulative stellar mass formed as a function of time. In the standard, core-collapse supernova feedback case, we see successive star formation epochs throughout the simulation. The core-collapse supernova feedback occurs $5$ Myr following a star formation epoch, temporarily reheating ambient gas in the region. Once this supernova event ends, the gas cools again and forms stars. In contrast, in the case of high $\sigma_{nX}$, dark baryonic feedback  causes sufficient explosions on short timescales that the galaxy has only one epoch of star formation before gas is ejected from the galaxy, and star formation ceases. At intermediate $\sigma_{nX}$, dark baryonic feedback delays subsequent star formation epochs, but does not shut them off completely. 

This interplay between dark baryonic feedback and star formation rate indicates that the effect of dark baryonic feedback on galaxy properties is non-linear across dark matter cross-sections. At low cross-sections, dark baryonic feedback occurs on too long a timescale to have a sizeable effect on star formation. While the lowest cross-section we have modelled in this study ($10^{-46}~{\rm cm^2}$) creates white dwarf explosions over Gyr timescales and has some effect on galaxy properties, a cross-section of order $10^{-47}~{\rm cm^2}$ would result in white dwarf explosions over 10 Gyr timescales, which would not occur over the timescale of the simulation. For higher dark matter-nucleon cross-sections, dark matter capture in white dwarfs is sufficiently efficient that white dwarfs of all masses can explode in quick succession, shutting off future star formation and baryonic feedback. At intermediate cross-sections, a balance is struck and dark baryonic feedback is efficient enough to slow gas cooling and star formation epochs, but still low enough to allow further stars to be formed and thus, dark baryonic feedback and standard feedback to be sustained in the galaxy.

We can see this in Figure \ref{fig:analyticandsim}, which shows the dark matter density profile of the galaxy after $2.5$ Gyr of evolution. All simulations show stellar feedback-induced flattening of the density profile, but there is a difference in efficiency. The low cross-section dark baryonic feedback cases are comparable to the core-collapse-only case, as the white dwarf explosions make up only a small fraction of the total feedback. At high cross-sections, however, dark baryonic feedback has a certain different influence on the central distribution of dark matter: its early shut-off of star formation results in a dark matter central distribution that is not markedly different from the normal feedback case. Here, an early sizable injection of dark baryonic feedback cores the galaxy, but also shuts off future star formation and core-collapse feedback. The simulation which results in the most coring is for an intermediate dark matter cross-section, where dark baryonic feedback continues throughout the simulation alongside moderate star formation and core-collapse feedback.

\edits{We note that the high Type-Ia supernova rates in the $\sigma_{nX} = 10^{-41}$ cm$^2$ simulations may be higher than observed Type-Ia rates per solar mass in dwarf galaxies, possibly rendering these models unrealistic. However, this conclusion will depend on the observability of Type-Ia supernovae at the center of dwarf galaxies, with possibly a higher optical extinction and corresponding lower observability for supernova events, since it is the central region where dark matter would preferentially ignite Type-Ia supernovae. Hence we include the result of these simulations, which also show the effects of dark baryonic feedback due to Type-Ia explosions with short explosion times, as compared to models that yield Type-Ias over a larger timescale. We also briefly comment on the effect of dark matter-induced Type-Ia supernovae on the global Type-Ia rate, which has been the object of many observational efforts as a probe of potential Type-Ia progenitors \cite{Maoz:2011iv,Phillips:2013doa,Chakrabarti_2018,brown_relative_2019}. In this work we refrain from directly performing an estimate of the global Type-Ia rate to constrain the dark matter parameter space in this work, as it would be reliant on modelling of star formation histories across varied galaxy masses, and the observability of Type-Ia supernovae in the nuclear centers of galaxies, as just discussed, but this could be a fruitful line of inquiry in future work.}

\section{Conclusion}

We have explored how dark matter-induced baryonic feedback via Type Ia supernovae affects galactic structure. Dark baryonic feedback's strong dependence on dark matter halo properties in the star-forming regions, as well as the interplay between dark baryonic feedback and standard baryonic effects, indicate that it should be simulated across a wider range of galaxies to determine its expected effects. The isolated dwarf galaxy we have simulated is small, with a low dark matter velocity dispersion, allowing quicker dark matter capture as well as more efficient gas outflows from the galaxy - an environment where suppressing star formation is much easier.

The impact of simulation resolution must also be pointed out. The resolution is determined by the mass ascribed to each test particle representing dark matter, gas, or stellar matter. In these low-resolution simulations, gas particles and therefore the star particles they form are on the order of $10^3$ solar masses. \edits{At higher resolution, we might expect to better model star formation in regions with lower gas density, such as regions outside of the galactic centre.} Appropriately scaling resolution will be important in future work -- in this proof-of-concept paper, we have kept resolution constant across all simulations to draw fair comparisons between different dark matter models. 

We have established how dark baryonic feedback can affect a dwarf mass galaxy, including quenching of star formation, which has been observed in dwarf galaxies \cite{Sharma2022}. However, there are potentially distinct properties of dark baryonic feedback that should be explored by extending this work to larger galaxies. In larger galaxies, the difference in dark baryonic feedback rates between regions of different dark matter density might be more easily observed. The impact of this effect on creating baryonic outflows from the center of very massive galaxies should be particularly examined and compared to active galactic nucleus feedback currently assumed to occur in these regions.  We leave this and other aspects of dark matter-induced baryonic feedback to future work.

\acknowledgments{We are grateful to Melissa Diamond, Rebecca Leane, Nirmal Raj, Volker Springel, and Haibo Yu for useful correspondence and discussions. This work was supported by the Natural Sciences and Engineering
Research Council of Canada (NSERC) and the Canada
First Research Excellence Fund through the Arthur B.
McDonald Canadian Astroparticle Physics Research
Institute. The work of HA was supported in part by the National Key R\&D Program of China under Grants No. 2021YFC2203100 and No. 2017YFA0402204, the NSFC under Grant No. 11975134, and the Tsinghua University Dushi Program No. 53120200422. JFA is supported in part by
the U.S. Department of Energy under Contract DE-AC02-76SF00515.}

\bibliographystyle{JHEP.bst}
\bibliography{blank}

\onecolumngrid
\newpage
\appendix



\section{Multiscatter Capture of Dark Matter in White Dwarfs}
\label{app:multiscatter}

First we review particle dark matter capture in white dwarfs. A dark matter particle transiting the star will become gravitationally bound to it ($i.e.$ captured) when it loses its halo kinetic energy $m_X v_{\rm halo}^2/2$ through scatterings against the stellar constituents. For heavy dark matter particles with masses $m_X \gtrsim 10^6 \ \rm GeV$, multiple scatters are required for this, since in the limit $m_X \gg m_N$, where $m_N \sim \mathcal{O}(10 \ {\rm GeV})$ is the mass of white dwarf nuclei, the energy transfer is proportional to the ratio $m_N/m_X$, and thus only a small fraction of the dark matter energy is lost per scatter. 

In this multiscatter capture limit, the total rate at which dark matter is captured has been analyzed in \cite{Bramante:2017xlb}. The total capture rate is expressed as
\begin{equation}
    C_X = \sum_{N=1}^\infty C_N,
    \label{eq:app-a-series}
\end{equation}
where each term $C_N$ is the capture rate for dark matter particles that undergo exactly $N$ scatters before being captured, measured in $\rm particles \times time^{-1}$. Each term is given by the expression,
\begin{equation}
    C_N = \pi{R_*^{2}}p_{N}(\tau)\sqrt{\frac{6}{\pi}}\frac{n_{X}}{v_{\rm halo}}\left[(2v_{\rm halo}^{2}+3v_{\rm esc}^{2})-(2v_{\rm halo}^{2}+3v_{N}^{2})\exp{\left(-\frac{3(v_{N}^{2}-v_{\rm esc}^{2})}{2v_{\rm halo}^{2}}\right)}\right]
    \label{eq:app-a-terms},
\end{equation}
where $v_{\rm esc}$ is the escape speed of dark matter from the surface of the white dwarf, $v_N \equiv v_{esc}(1-\beta_{+}/2)^{-N/2}$ with $\beta_{+}=4m_{X}m_{N}/(m_{X}+m_{N})^{2}$ for nucleus mass $m_N$, and the probability for capture during transit is
\begin{equation}
    p_N(\tau) = \frac{2}{N!} \int_{0}^{1} \left(\cos\theta\right)^{N+1} \tau^N \exp\left(-\tau \cos\theta\right)\ d\left(\cos\theta\right).
    \label{eq:app-a-probs}
\end{equation}
Here, $\tau = 3 \sigma_{NX} M_* / 2 \pi R_*^2 m_N$ is the stellar optical depth, which is the ratio between the dark matter-nucleus cross-section $\sigma_{NX}$ and the saturation cross-section for which a dark matter particle scatters once on average while transiting a distance $2R_*$ in the stellar interior. This cross-section can be parametrized as
\begin{equation}
    \sigma_{NX} = A^2 \left(\frac{\mu_{NX}}{\mu_{nX}}\right)^2 |F_{\rm helm}\left(\bar{q}\right)|^2 \sigma_{nX},
    \label{eq:app-a-cs}
\end{equation}
where $A \sim 12 - 16$ is the mass number of the white dwarf nuclei, $\mu_{NX}$ ($\mu_{nX}$) is the dark matter-nucleus (dark matter-nucleon) reduced mass, and $\sigma_{nX}$ is dark matter-nucleon cross-section. The function $|F_{\rm helm}\left(\bar{q}\right)|^2 \sim 0.5$ is the Helm form factor evaluated at the average momentum transfer per scatter $\bar{q}$ per scatter, \edits{of order $\sim 0.5$} for dark matter scattering at $v \sim v_{\rm esc} \sim 10^{-2}$ in white dwarfs. The factor $p_N(\tau)$ is the probability of a dark matter particle to scatter $N$ times, and it is a Poissonian distribution integrated over all possible entry angles $\theta$. Eq.~\eqref{eq:app-a-cs} assumes that the energy loss per scatter is uniformly distributed over the interval $0 < E < \beta_{+} E_0$ and $E_0$ is the kinetic energy of the dark matter particle prior to the scattering event in the stellar rest frame. These equations were used to derive the fit capture rate given below, see \cite{Bramante:2017xlb,Acevedo:2019gre} for further details on the derivation. Finally, we remark that for dark matter masses $m_X \lesssim \rm PeV$, one recovers the single-scatter capture treatment, given by the first term $C_1$ in Eq.~\eqref{eq:app-a-terms}. This multi-scatter capture regime was first analyzed in \cite{Bramante:2017xlb}, to which we refer for its technical details. Here, we are primarily concerned with providing a useful fit for the total mass capture rate in this limit \cite{Acevedo:2019gre},
    \begin{align}
         C_X \simeq \left(\frac{\rho_X}{0.4 \ {\rm GeV cm^{-3}}}\right) \left(\frac{v_X}{10^{-3}}\right)^{-1} \min \left[10^{27} \ {\rm GeV \ s^{-1}}, 10^{25} \ {\rm GeV \ s^{-1}} \left(\frac{m_X}{100 \ \rm PeV}\right)\left(\frac{\sigma_{nX}}{10^{-40} \ \rm cm^2}\right)\right]~.
         \label{eq:cx}
     \end{align}

\section{Type-Ia Supernovae Ignition through Asymmetric Dark Matter Capture}
\label{app:ignition}

\begin{figure}[h!]
\centering 
\includegraphics[width = \textwidth]{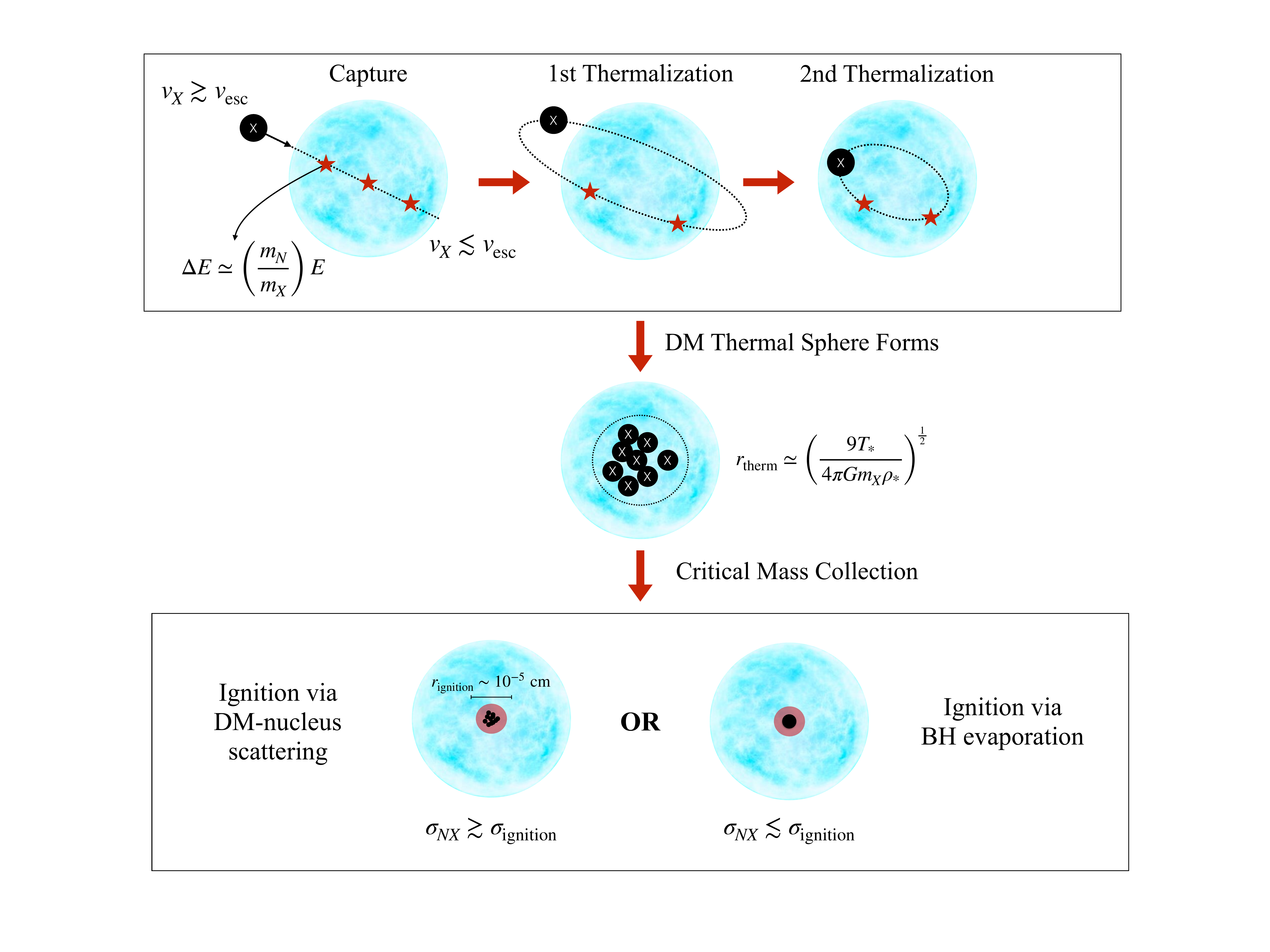} 
\caption{Asymmetric dark matter ignition of Type-Ia supernovae in massive white dwarfs. The upper panel shows the basic process through which dark matter settles at the core of the white dwarf. Each red star represents a scattering event whereby a dark matter particle loses a small fraction of its kinetic energy in the limit $m_N \ll m_X$, with $m_N \sim \mathcal{O}(15 \ \rm GeV)$. The central panel shows the dark matter after it settles in a compact configuration at the core. Since dark matter is asymmetric, it accumulates in large amounts, eventually reaching a gravitational instability and collapsing. The lower panel shows how this collapse can spark a Type-Ia supernova: if the cross-section is above a certain threshold $\sigma_{\rm ignition}$, the scattering of dark matter particles with stellar nuclei as they collapse is enough to trigger ignition. If that is not the case, ignition can still proceed in certain regions of parameter space if the collapsed dark matter forms a light black hole that subsequently evaporates to Hawking radiation. The hadronic showers produced by the radiated color-charged particles, as they scatter against white dwarf material, are sufficient to ignite the small trigger region denoted by $r_{\rm ignition}$. See text and Refs.~\cite{Bramante:2015cua,Acevedo:2019gre,Janish:2019nkk} for more details.}
\label{fig:snia-shematic}
\end{figure}

Figure~\ref{fig:snia-shematic} depicts how this ignition mechanism proceeds: some dark matter particles passing through the white dwarf will become gravitationally bound to it, due to the energy lost as they scatter against stellar constituents. These captured dark matter particles follow closed orbits that intersect with the white dwarf. The orbits progressively shrink with each crossing as the dark matter particles lose more energy, eventually becoming fully contained within the white dwarf volume. Further loss of energy leads to the dark matter \edits{forming a thermalized sphere within the white dwarf.}

If the dark matter is asymmetric or otherwise weakly-annihilating, then large amounts accumulate over time \edits{ compared to annihilating dark matter.} The dark matter core's mass grows over time as more dark matter particles are captured and thermalized, however its size is fixed by the mass of the dark matter and the white dwarf temperature. Thus, the dark matter density steadily increases until eventually the sphere achieves self-gravitation. When this occurs, it is energetically favourable for the sphere to shrink in size. As we detail below, the sphere also develops a gravitational instability, prompting its collapse. 

As the sphere collapses, the shedding of its gravitational energy through repeated scatterings of the dark matter particles against nuclei heats a small central region of the white dwarf.
If the dark matter-nucleus cross-section is large enough, this energy transfer rate is sufficient to cause a thermonuclear runaway in the star, leading to a Type-Ia supernova. This is depicted in the lower-left side of Fig.~\ref{fig:snia-shematic}. Alternatively, as argued in \cite{Acevedo:2019gre,Janish:2019nkk}, for higher mass dark matter or lower cross-sections (not considered in the main text of this paper), ignition can still occur even when the cross-section between the dark matter and the stellar constituents is inadequate to prompt ignition during the collapse of the dark matter core, $i.e$ the white dwarf material effectively dissipates the excess heat. In this case, the dark matter sphere collapses without igniting the star, and forms a black hole at its center. The evolution of such a black hole is determined by its initial mass: if it is sufficiently heavy, it will grow by accretion of stellar material and newly-captured dark matter. This ultimately leads to the implosion of the white dwarf, without producing a Type-Ia supernova. This scenario is not relevant for the purposes of this work. By contrast, if the black hole is light enough, it will evaporate to Hawking radiation. As it was shown in \cite{Acevedo:2019gre}, the color-charged particles emitted from the horizon will scatter against white dwarf nuclei within a short mean free path, producing hadronic showers that distribute most of the energy homogeneously within the ignition region. This alternative ignition channel is depicted on the lower-right side of Fig.~\ref{fig:snia-shematic}.

There are three physical timescales that determine how the time it takes for a massive white dwarf to explode through dark matter accumulation. These are: 1) the time it takes for the captured dark matter particles to thermalize at the center, 2) the time it takes to capture the critical mass for dark matter, and 3) the time it takes for the dark matter core to collapse. The upper and middle panels of Fig.~\ref{fig:snia-shematic} illustrate these steps. The above capture rate partially determines the ignition timescale, since it determines for a given cross-section and dark matter mass the time it takes for the dark matter sphere to become unstable and collapse. Below, we proceed to detail the additional timescales involved with the white dwarf ignition process.

\begin{figure}[h!]
\centering 
\includegraphics[width = 0.7\textwidth]{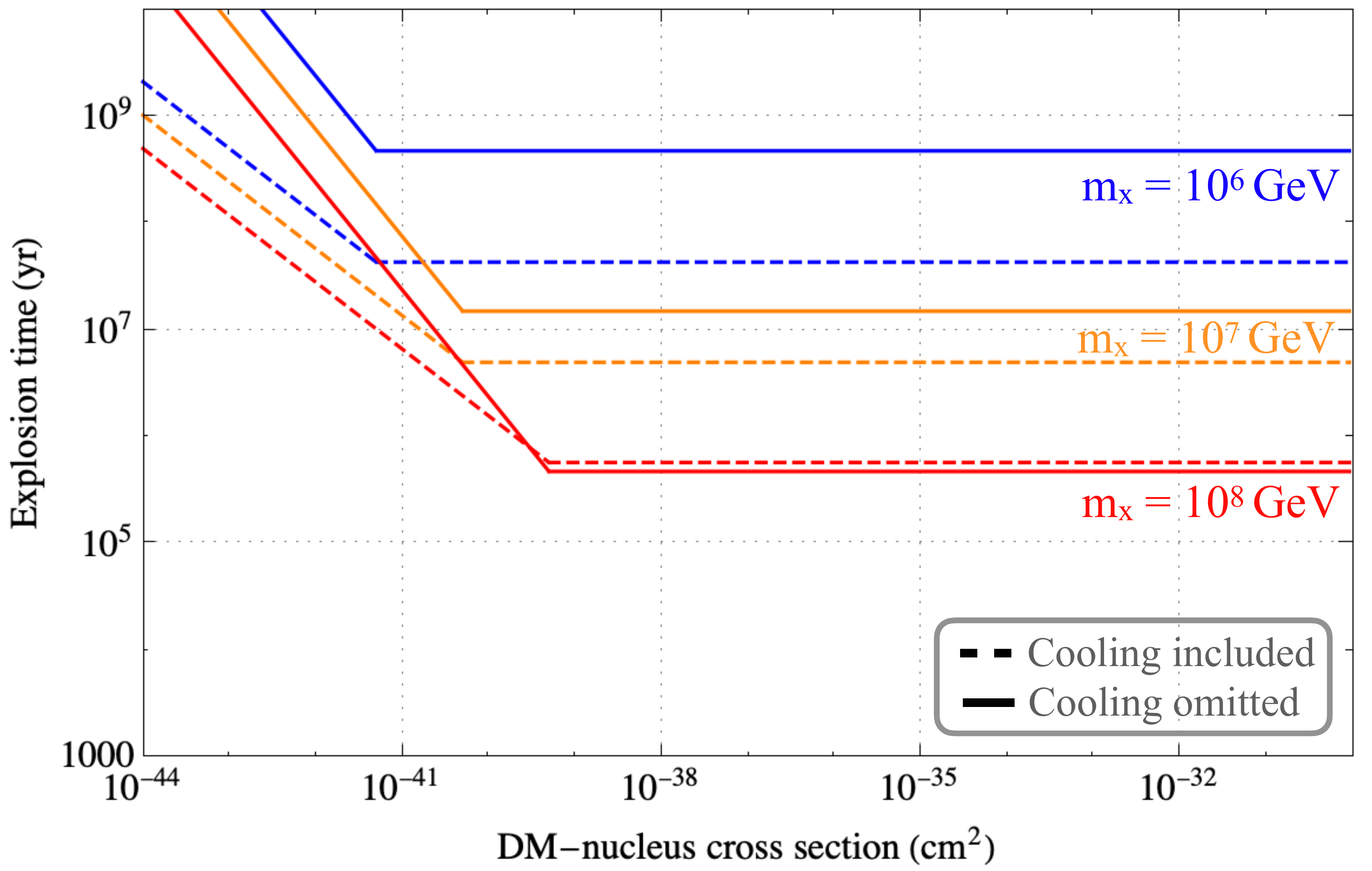} 
\caption{Type Ia supernova ignition time as a function of the dark matter-nucleus cross-section, for various dark matter masses as specified. The dashed lines correspond to the ignition time accounting for white dwarf cooling using a simple Mestel's law $T(t) \propto t^{-2/5}$, which subsequently lowers the amount of dark matter needed to form a self-gravitating sphere at the center of the white dwarf. The background dark matter density and velocity assumed corresponds to those given in Eq.~\eqref{eq:cx}.}
\label{fig:explotimes1}
\end{figure}

\begin{enumerate}
     \item \textit{Thermalization:} Once the dark matter particles are captured, they must further lose energy to settle at the center of the white dwarf. As discussed in \cite{Kouvaris:2010jy,Acevedo:2019gre}, the dark matter thermalization proceeds in two steps: first, a captured dark matter particle orbits around the white dwarf, crossing it multiple times, and losing a fraction of its energy in each passage. The energy loss rate The energy loss rate in this first thermalization stage can be obtained by estimating the average energy loss rate per crossing, and dividing this quantity by the orbital period, the latter being a function of the energy itself. Assuming an approximately uniform white dwarf density profile, the first thermalization timescale is \cite{Acevedo:2019gre},    
     \begin{equation}
         t_1^{\rm th} \simeq 10^{-4} \ {\rm yrs} \left(\frac{m_X}{\rm PeV}\right) \left(\frac{\sigma_{nX}}{10^{-40} \ \rm cm^2}\right)^{-1} \left(\frac{M_*}{1.4 M_\odot}\right)^{-\frac{3}{2}} \left(\frac{R_*}{2500 \ \rm km}\right)^{\frac{7}{2}}~.
     \end{equation}     
     Once the orbits of dark matter particles are fully contained within the white dwarf, the so-called second thermalization stage commences. For heavy dark matter particles, their final thermal velocity $v_{\rm th} \simeq (T_*/m_X)^{1/2}$ is much smaller than the thermal speed of the white dwarf nuclei. Furthermore, as the inverse momentum transfer becomes longer than the interparticle separation as this thermal velocity is approached, coherent nuclear scattering becomes suppressed compared to the excitation of phonons. In \cite{Acevedo:2019gre}, this was accounted for by introducing an additional form factor $1 - \exp\left(W(q)\right)$ in the cross-section, where $W(q) \sim T_* q^2 / m_N \omega_p$ is the Debye-Waller factor of the lattice in the long-wavelength limit. Because of this, the second thermalization timescale computed is far longer than the previous one \cite{Acevedo:2019gre}, 
     \begin{equation}
         t_2^{\rm th} \simeq {\frac{m_{X}E_{sup}\sqrt{m_{a}}}{3T^{3/2}\rho_{wd}A^{4}\sigma_{nX}}} \nonumber \\  \approx{20 \ {\rm yrs} \left(\frac{10^{-40} \ {\rm cm^{2}}}{\sigma_{nX}}\right)\left(\frac{m_{X}}{10^{6} \ \gev}\right)^{2}\left(\frac{10^{7} \ \rm K}{T}\right)^{\frac{5}{2}}}~.
     \end{equation} 
     \item \textit{Self-gravitating mass:} The dark matter sphere can only collapse if it becomes self-gravitating, $i.e.$ a contraction leads to a reduction of its total gravitational potential energy. This condition is fulfilled when its density $\rho_X$ approximately matches the white dwarf background density $\rho_*$. The dark matter thermal sphere has a radius that can be estimated utilizing the virial theorem \cite{Acevedo:2020avd}
     \begin{equation}
         r_{\rm th} \simeq \left(\frac{9 T_*}{4 \pi G \rho_* m_X}\right)^\frac{1}{2} \simeq 30 \ {\rm m} \left(\frac{m_X}{\rm PeV}\right)^{-\frac{1}{2}} \left(\frac{\rho_*}{10^9 \ \rm g \ cm^{-3}}\right)^{-\frac{1}{2}} \left(\frac{T_*}{10^7 \ \rm K}\right)^\frac{1}{2}
     \end{equation}
     
     Self-gravitation then occurs when $N_X r_{\rm th}^3 \sim \rho_*$, where $N_X$ is the number of captured and thermalized dark matter particles and $\rho_* \simeq 10^7$ - $10^9 \ \rm g \ cm^{-3}$ in the white dwarf mass range of interest. This condition sets a critical mass of dark matter to be captured
     
     \begin{equation}
         M_{\rm crit} \simeq 10^{44} \ {\rm GeV} \left(\frac{m_X}{\rm PeV}\right)^{-\frac{3}{2}} \left(\frac{\rho_*}{10^9 \ \rm g \ cm^{-3}}\right)^{-\frac{1}{2}} \left(\frac{T_*}{10^7 \ \rm K}\right)^{\frac{3}{2}},
         \label{eq:mcrit}
     \end{equation}
     
     in order for the dark matter sphere to be self-gravitating. As shown in \cite{Acevedo:2019gre}, once self-gravitation is attained, a simple Jeans instability criterion is also satisfied. Thus, the sphere will collapse once the critical mass above has been captured and thermalized.  The timescale for capturing the critical mass is thus,     
$     
        t_{\rm crit}\simeq \frac{M_{\rm crit}}{C_X},
$
where this is obtained from Eqs.~\eqref{eq:cx}~and~\eqref{eq:mcrit}. As discussed shortly, this will be the limiting timescale for white dwarfs to explode, and is the timescale shown in Fig.~\ref{fig:explotimes1}. Note that recently, Ref.~\cite{Acevedo:2023xnu} showed that capture rates in white dwarfs are smaller than previously expected due to quantum effects associated with the white dwarf microscopic structure as the object cools down. However, such analysis was limited to the capture in the optically-thin regime where dark matter particles are captured in about a single scatter per crossing. In the multi-scatter regime we consider here, we expect this capture rate suppression to weaken as the dark matter particles are slowed down within the white dwarf. Furthermore, we emphasize that dark baryonic feedback occurs for a wide range of masses and cross-sections that can easily accommodate such suppression in the capture rate, and thus the associated uncertainty in the timescale to accumulate $M_{\rm crit}$ in a white dwarf. 
    
    \item \textit{Collapse:} Once the dark matter sphere becomes gravitationally unstable, the timescale for its collapse is determined by how frequently the dark matter scatters against nuclei. The initial stage of collapse proceeds very slowly, as the velocity of the dark matter particles is near thermal. As detailed above, coherent nuclear scattering in this regime becomes suppressed compared to phonon excitations and energy loss proceeds very slowly. Once the dark matter sphere has collapsed to the point that dark matter particles move fast enough for DM-nucleus scattering to be the main energy loss channel, this process continues at a faster rate. This proceeds until dark matter-nucleus scattering becomes suppressed once again due to nuclear substructure effects at high momentum transfers, incorporated the Helm form factor. The timescale for this is approximately given by \cite{Acevedo:2019gre}
    \begin{equation}
        t_{\rm col} = 10 \ {\rm yr} \left(\frac{\sigma_{nX}}{10^{-40} \ \rm cm^2}\right)^{-1} \left(\frac{m_X}{10^6 \ \rm GeV}\right)^2 \left(\frac{T_*}{10^7 \ \rm K}\right)^{\frac{5}{2}}~.
    \end{equation}
\end{enumerate}
Once the dark matter sphere collapses, ignition will occur as long as certain conditions are met. These are: 1) the dark matter must heat the white dwarf faster than it can dissipate energy, and 2) a critical temperature of $T_{\rm crit} \simeq 10^{10} \ \rm K $ $(\sim \mev)$ must be achieved within a small central region. The internal heat transport of the white dwarf is dominated by electron conduction. The rate at which the white dwarf dissipates heat from a spherical region of radius $r$ is determined mainly by electron conduction $\dot{Q}_{\rm cond}$, detailed in \cite{Bramante:2015cua,Acevedo:2019gre}. On the other hand, the dark matter heating rate is roughly determined by the scattering frequency, $t_{NX}^{-1}$, the average energy exchange per scatter $\Delta{E}$ and the number of self-gravitating dark matter particles $N_{\rm sg}$ \cite{Acevedo:2019gre}
\begin{equation}
    \dot{Q}_{\rm DM} \simeq \frac{N_{sg} \ \Delta{E}}{t_{NX}}\approx{2 \times{10^{37}} \  { \rm \frac{GeV}{s}} \ \left(\frac{\sigma_{nX}}{10^{-40} \ \rm cm^{2}}\right) \left(\frac{10^{6} \ \gev}{m_{X}}\right)^{\frac{5}{2}}  \left(\frac{\rho_{wd}}{10^{9} ~{\rm g \ cm^{-3}}}\right)^{\frac{1}{2}} \left(\frac{T}{10^7 \ \rm K}\right)^{\frac{3}{2}}}~,
\end{equation}
where the rightmost expression is evaluated when the dark matter particles are moving at a speed $v_{\rm vir} \simeq 0.02 \ c$, corresponding to the maximum value before coherent enhancement in the scattering is lost.

The total explosion time since a white dwarf is formed is thus given by the sum above timescales. In practice, for the majority of the parameter space, only one of the above timescales dominates the explosion time. In this work, we have limited ourselves to dark matter mass and cross-section values where the capture of critical mass is what sets the explosion time, $i.e.$ $t_{\rm crit}$ as defined above. 
Furthermore, since white dwarfs within dark matter backgrounds of large density will explode within a brief period of time after formation, we can correct this estimate by accounting how the core temperature drops using a simple Mestel's law estimate \cite{1983bhwd.book.....S}, $T_*(t) \simeq T_0 (t/t_0)^{-2/5}$ where $T_0$ and $t_0$ denote the initial time and temperature at which white dwarf cooling is dominated by photon emission. We have also verified, using density profiles for various white dwarf progenitors simulated with MESA \cite{Mesa}, that dark matter capture during the main-sequence or asymptotic giant branch phase is negligible, since for the low nucleon cross-sections we consider only a small fraction of heavy dark matter is captured in main sequence stars. Indeed, using a single-scatter capture treatment, appropriate for small optical depths, we found that the white dwarf progenitor at most captures a fraction $\sim \mathcal{O}(10^{-2})$ of the dark matter required for ignition during its $10 - 50 \ \rm Myr$ lifetime. 

Finally, we also comment on the ignition conditions and the potential impact of the white dwarf composition. The ignition critical temperature and trigger mass outlined above have been numerically computed for either pure $^{12}$C, or $^{12}$C and $^{16}$O (C/O) white dwarfs in various ratios, with central densities roughly corresponding to the range $1.1 - 1.4M_\odot$. Although massive white dwarfs in this range are expected to have cores predominantly made of $^{16}$O and $^{22}$Ne (O/Ne), there are several studies \cite{2017A&A...602A..16T,Maoz:2018epf,2020A&A...636A..31T} showing that a significant fraction of these objects are C/O white dwarfs formed in binary mergers \cite{Yoon:2007pw,LorenAguilar:2009cv}. Furthermore, single-evolution channels may also produce C/O white dwarfs, including reduced mass-loss rates \cite{2019NatAs...3..408D,2021A&A...646A..30A} or enhanced rotation \cite{1996ApJ...472..783D,2021A&A...646A..30A} of the progenitor during the asymptotic giant branch phase.


In Figure \ref{fig:semianalytic} we show some simplified exploration of baryonic feedback in a semi-analytic model galaxy, subject to certain limits, as an extension of Section 3 of the main text. 

Whether the potential increases to its original value or only fractionally depends on the intensity of the feedback. In this work, we consider a gradual recollapse process in which the potential is increased over a timescale of $\sim 50$ Myr. We note that one of the variables governing the explosion time of a white dwarf due to dark matter capture is the white dwarf mass, with higher-mass white dwarfs exploding in a shorter time \cite{Bramante:2015cua,Acevedo:2019gre}. A single epoch of star formation therefore leads to multiple successive supernova events at different timescales.

To provide some approximate modeling of late-stage dark baryonic feedback processes in a galaxy, we consider a population of dark matter particles, distributed according to a simplified halo profile near a galactic centre. For simplicity we use a power-law gravitational potential in the interior region given by $\rho(r) = \rho_0 r^{-2}$, to illustrate baryonic feedback effects in a regime with a cuspy internal profile. We use a code that explicitly calculates the position and acceleration of the dark matter particles to determine their orbits following changes in the gravitational potential, which we model as
\begin{equation}
    \Delta V(t) = - \kappa_i V_0  \left( 1- \left( 1 - \exp\left[ -\frac{t}{t_{\rm recollapse}} \right] \right) \right)
\end{equation}
where $\kappa_i$ is the factor by which the potential is decreased after a supernova explosion. The exponential term implements the gradual recollapse of the potential after the explosion time $t=0$, for a recollapse time $t_{\rm recollapse} \approx 50$ Myr, which we assume is much shorter than the time between explosions.

To illustrate a qualitative difference between standard vs.~dark matter-induced baryonic feedback, in Figure~\ref{fig:semianalytic} we show how assuming a fixed values for $\kappa_i$($=0.1,0.5$) differs from assuming that $\kappa_i$ decreases with each explosive epoch as $\kappa_i=\kappa_1/i$, for epochs $i=1,2,...8$. This mimics an effect of dark baryonic feedback, which will tend to decrease the galactic center dark matter density and star formation rate, thereby decreasing future supernova rates and also dark baryonic feedback itself. 

\clearpage

\section{Additional Simulation Plots}
\label{app:simplots}

\subsubsection{Supernova Locations}
\begin{figure*}[th!]
    \centering
    \includegraphics[width=\linewidth]{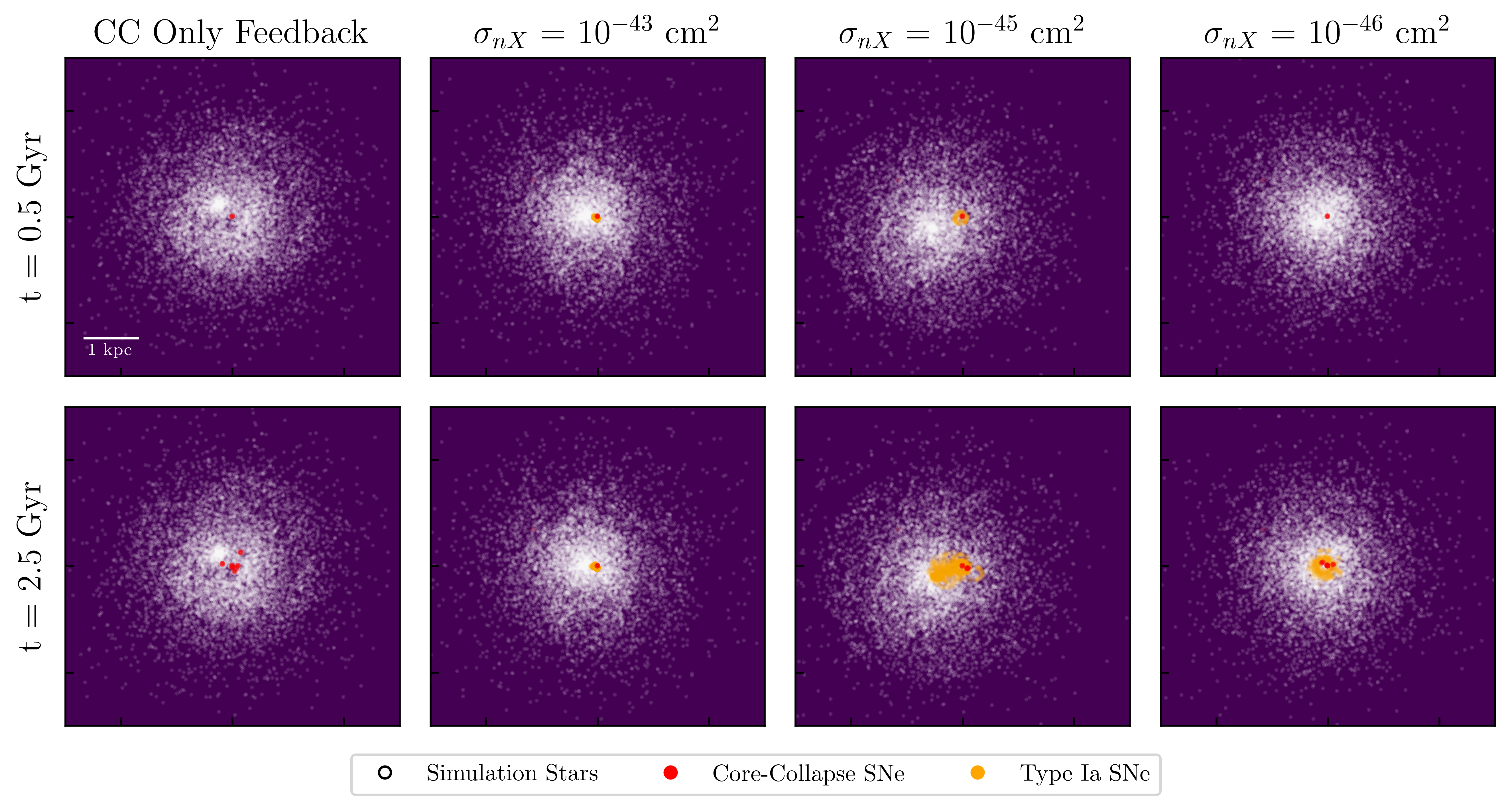}
    \caption{Star particle and supernova locations at early (t = 0.5 Gyr) and late (t = 2.5 Gyr)  simulation times,  for a core-collapse-only feedback scenario as well as dark baryonic feedback simulations for $10^7$ GeV dark matter and $\sigma_{nX}$ as indicated. Red points show locations of core-collapse supernova explosions up until the time indicated, and orange points show the same for  dark matter-induced Type Ia supernovae. It is important to note that many supernovae can explode from a single star particle, so single points indicating core-collapse or Type Ia supernovae could represent multiple supernovae. We also note that the star particles plotted in white include pre-existing stars from the galaxy initial conditions, and that pre-existing stars were not included in the core collapse or Type Ia feedback protocols.}
    \label{fig:supernovaloc}
\end{figure*}

In Fig. \ref{fig:supernovaloc} we show galaxies simulated in this study, with the historic locations of dark matter induced Type Ia and core collapse supernovae indicated at two timesteps during the evolution of the galaxies.

\clearpage

\subsubsection{Final dark matter distributions}

\begin{figure*}[th!]
    \centering
    \includegraphics[width=\linewidth]{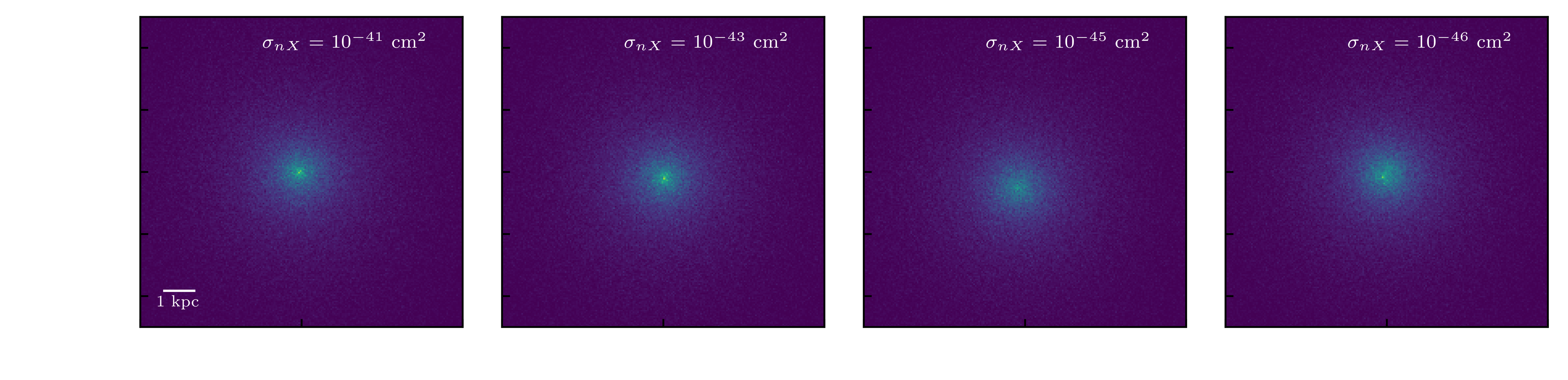}
    \caption{Final distribution ($t=2.5$ Gyr) of dark matter particles in simulated galaxies for $\sigma_{nX}$ indicated.}
    \label{fig:densitysnaps}
\end{figure*}

In Fig. \ref{fig:densitysnaps} we show the projected dark matter density distribution at the end of each galaxy simulation ($t=2.5$ Gyr), comparing dark baryonic feedback simulations for $10^7$ GeV dark matter and $\sigma_{nX}$ as indicated.
 
\subsubsection{Gas Density Plots}

\begin{figure*}[th!]
    \centering
    \includegraphics[width=.95\linewidth]{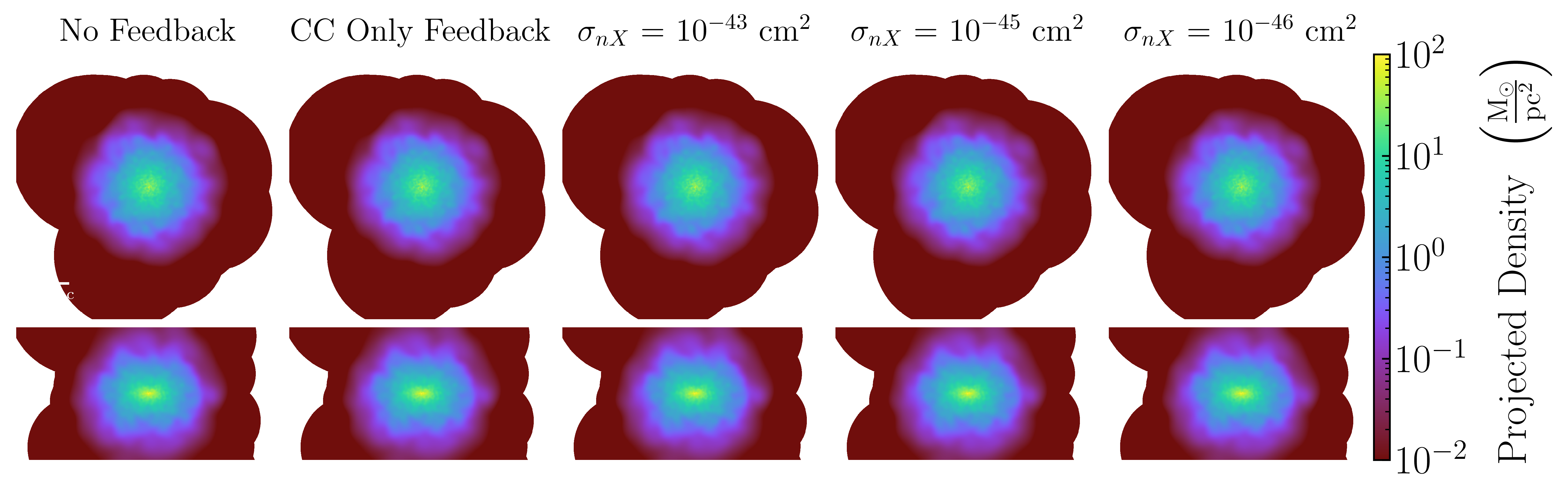}
    \caption{Gas density distribution in simulated galaxies from a face-on (top) and edge-on (bottom) view at timestep t = 0 Gyr.}
    \label{fig:gas0}
\end{figure*}

\begin{figure*}[th!]
    \centering
    \includegraphics[width=.95\linewidth]{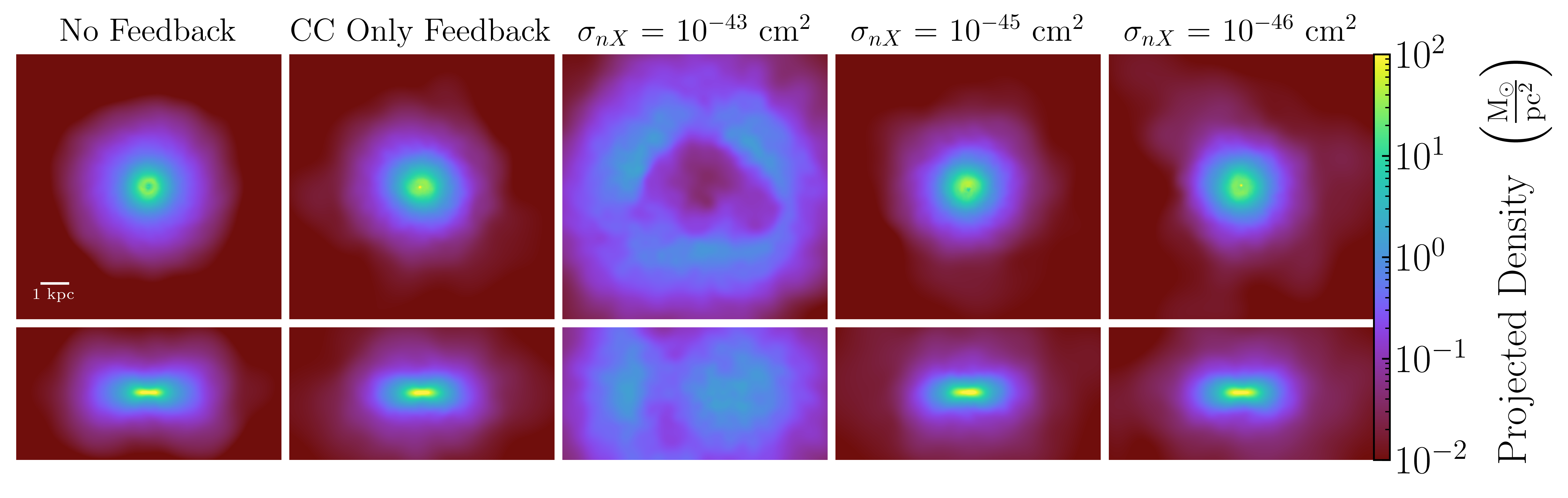}
    \caption{Same as Figure \ref{fig:gas0}, but for t = 0.5 Gyr.}
    \label{fig:gas50}
\end{figure*}

\begin{figure*}[th!]
    \centering
    \includegraphics[width=.95\linewidth]{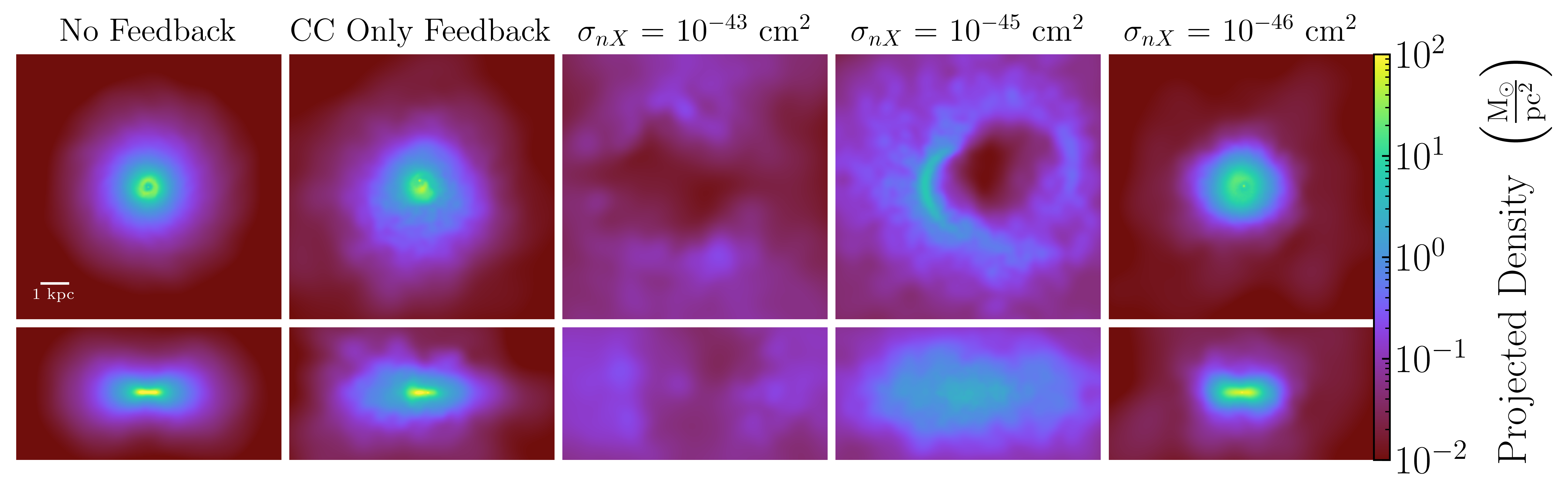}
    \caption{Same as Figure \ref{fig:gas0}, but for t = 1.0 Gyr.}
    \label{fig:gas100}
\end{figure*}

\begin{figure*}[th!]
    \centering
    \includegraphics[width=.95\linewidth]{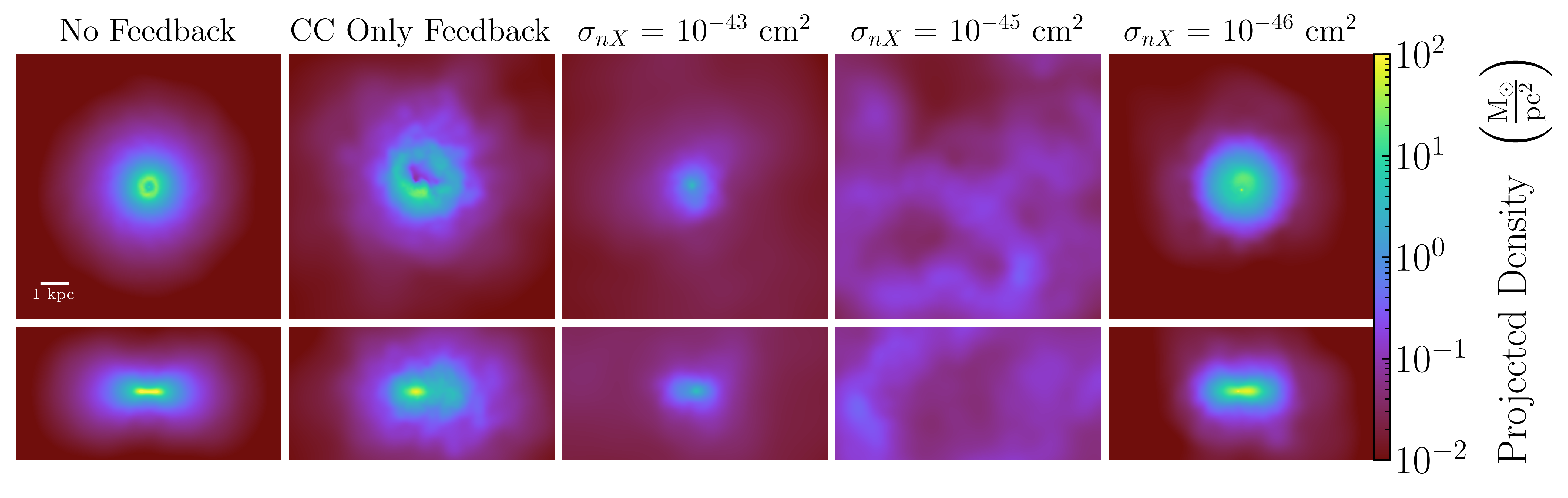}
    \caption{Same as Figure \ref{fig:gas0}, but for t = 1.5 Gyr.}
    \label{fig:gas150}
\end{figure*}

\begin{figure*}[th!]
    \centering
    \includegraphics[width=.95\linewidth]{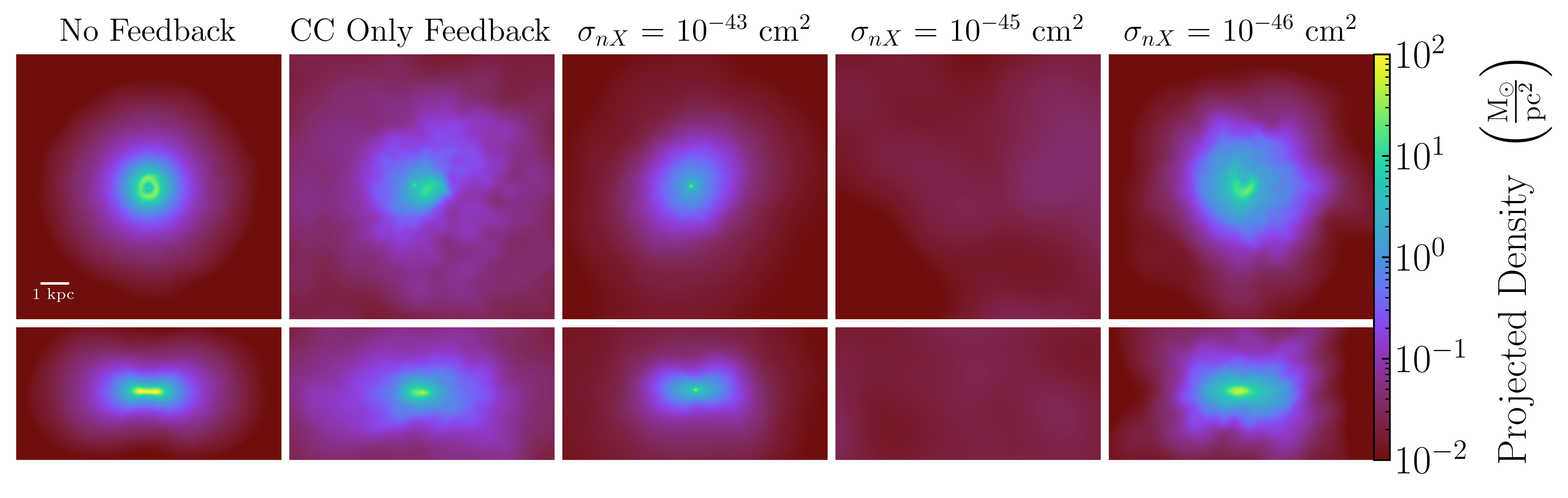}
    \caption{Same as Figure \ref{fig:gas0}, but for t = 2.0 Gyr.}
    \label{fig:gas200}
\end{figure*}

\begin{figure*}[th!]
    \centering
    \includegraphics[width=.95\linewidth]{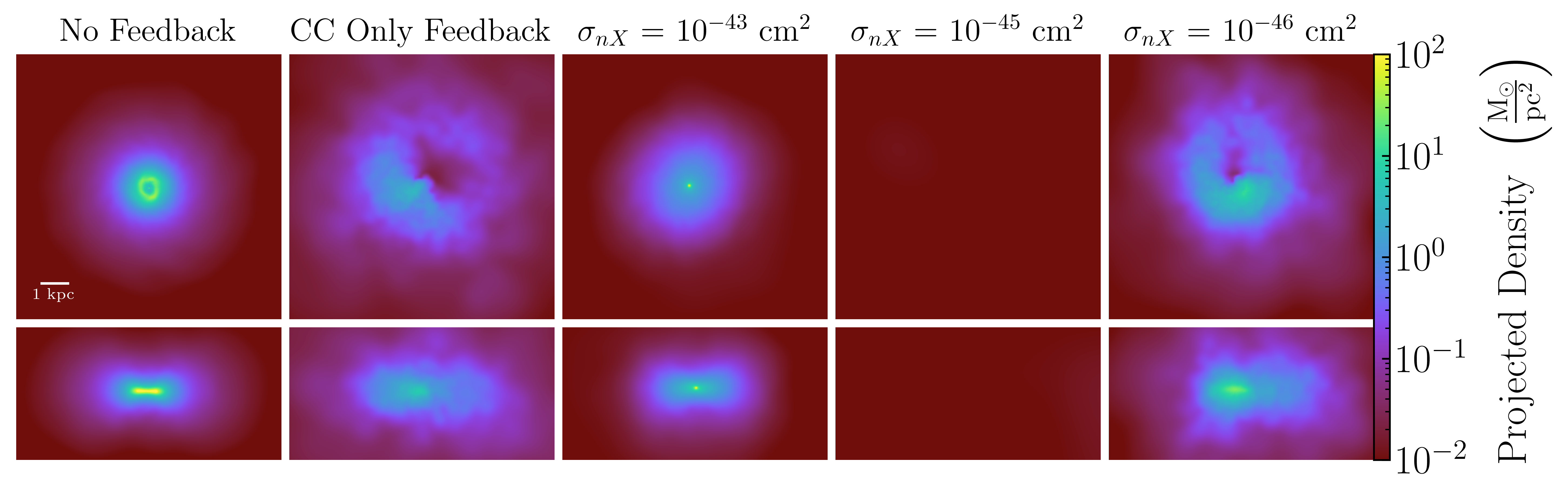}
    \caption{Same as Figure \ref{fig:gas0}, but for t = 2.5 Gyr.}
    \label{fig:gas250}
\end{figure*}

In Figs. \ref{fig:gas0}-\ref{fig:gas250}, we show five snapshots of the projected gas density at different timesteps of isolated galaxy simulations. At each timestep we compare a simulation without any feedback, a core-collapse-only feedback scenario, as well as dark baryonic feedback simulations for $10^7$ GeV dark matter and $\sigma_{nX}$ as indicated.

\clearpage

\clearpage

\end{document}